\documentclass[]{interact}

\usepackage{epstopdf}% To incorporate .eps illustrations using PDFLaTeX, etc.
\usepackage[caption=false]{subfig}% Support for small, `sub' figures and tables

\usepackage[numbers,sort&compress]{natbib}% Citation support using natbib.sty
\usepackage{amsmath}
\usepackage{mathtools}
\usepackage{url}
\usepackage{hyperref}
\bibpunct[, ]{[}{]}{,}{n}{,}{,}% Citation support using natbib.sty
\renewcommand\bibfont{\fontsize{10}{12}\selectfont}% Bibliography support using natbib.sty
\def\NAT@def@citea{\def\@citea{\NAT@separator}}% Suppress spaces between citations using natbib.sty
\makeatother% @ becomes a symbol again

\theoremstyle{plain}% Theorem-like structures provided by amsthm.sty

\theoremstyle{definition}

\theoremstyle{remark}

\begin{document}

%\articletype{ARTICLE TEMPLATE}% Specify the article type or omit as appropriate

\title{Interstellar Objects}

\author{
\name{Darryl ~Z. Seligman\textsuperscript{a}\thanks{CONTACT D.~Z. Seligman Email: dzs9@cornell.edu} and Amaya Moro-Mart\'in \textsuperscript{b}\textsuperscript{c}}
\affil{\textsuperscript{a}Department of Astronomy and Carl Sagan Institute, Cornell University, 122 Sciences Drive, Ithaca, NY, 14853, USA; \textsuperscript{b}Space Telescope Science Institute, 3700 San Mart\'in Dr., Baltimore, MD 21218; \textsuperscript{c} Department of Physics and Astronomy, Johns Hopkins University, Baltimore, MD 21218, USA}
}

\maketitle

\begin{abstract}
Since 2017, two macroscopic interstellar objects have been discovered in the inner Solar System, both of which are distinct in nature. The first interstellar object, 1I/`Oumuamua, passed within $\sim63$ lunar distances of the Earth, appeared asteroidal lacking detectable levels of gas or dust loss, yet exhibited a nongravitational acceleration.  1I/`Oumuamua's brief visit left open questions regarding its provenance which has given rise to many theoretical hypotheses, including an icy comet lacking a dust coma, an elongated fragment of a planet or planetesimal that was tidally disrupted, and an ultra-porous fractal aggregate. The second interstellar object, 2I/Borisov, was distinct from 1I/`Oumuamua in terms of its bulk physical properties and displayed a definitive cometary tail. We review the discoveries of these objects, the subsequent observations and characterizations, and the theoretical hypotheses regarding their origins. We describe 1I/`Oumuamua and 2I/Borisov in the context of active asteroids and comets in the Solar System.  The discovery of these two objects implies a galactic-wide population of $\sim10^{26}$ similar bodies. Forthcoming observatories should detect many more interstellar planetesimals which may offer new insights into how  planetary formation processes vary throughout the Galaxy.
\end{abstract}

\begin{keywords}
Interstellar objects; comets; asteroids; meteoroids; planetary systems: formation, protoplanetary disks, circumstellar disks; dynamical evolution and stability; Solar System: formation; Oort cloud; galaxy: local interstellar matter.
\end{keywords}
\section{Comets in the Solar System}

\subsection{Historical Perspective}

  For hundreds of years prior to the discovery of 1I/`Oumuamua and 2I/Borisov, astronomers had predicted that the Galaxy would be filled with icy bodies ejected from extrasolar planetary systems. Therefore, the existence of small bodies of extrasolar provenance observed passing through our own Solar System on hyperbolic trajectories was not inherently surprising. A full detailed account of the historical perspective on cometary science can be found in \cite{Yeomans1991} including a discussion of interstellar comets. 

As early as 1705, Edmond Halley considered the hypothesis that some   comets arrived from interstellar space \citep{Halley1705}. Halley computed the orbits of 24 comets and demonstrated that none of them had distinctly hyperbolic trajectories --- outlined in the Table captioned \textit{The astronomical elements of the motions in a parabolick orb of all the Comets that have been hitherto duly observ'd}. He therefore concluded that the population of detected comets at the time had all formed within the Solar System. Immanuel Kant further postulated that comets formed from diffuse nebular material at large  heliocentric  distances but were still gravitationally bound to the Sun \citep{Kant1755}. 

 On the other hand, the Italian astronomer Giuseppe (translated in English to Joseph) Lagrange  argued that the high eccentricities and inclinations of cometary orbits was incompatabile with the  nebular hypothesis. Although he believed that comets formed within the Solar System, he showed that their orbits could be produced by catastrophic disruption events from impacts with planets. Lagrange argued that his calculations, in combination with the nebular hypothesis, explained the entire census of celestial objects within the Solar System \citep{lagrange1812}.

In 1812, William Herschel obtained detailed observations of two comets, both of which attained different brightness levels despite similar perihelia distances.  He outlined these findings in two papers published in 1812: “Observations of a Comet, with Remarks on the Construction of Its Different Parts” and “Observations of a Second Comet, with Remarks on Its Construction" \citep{Herschel1812, Herschel1812b}. The brightness differences led him to speculate that some comets may originate beyond the Solar System, and that the variable activity levels could be attributed to the accumulation of matter in interstellar space: 

\textit{``Should the idea of age be rejected, we may indeed have recourse to another supposition, namely, that the present comet, since the time of some former perihelion passage, may have acquired an additional quantity (if I may so call it) of unperihelioned matter, by moving in a parabolical direction through the immensity of space, and passing through extensive strata of nebulosity."}

In 1813 Pierre-Simon de Laplace offered an alternative hypothesis to that of Lagrange \citep{Laplace1816}. He argued that the eccentricities and inclinations of comets could be explained if they originated from interstellar space, echoing  the hypothesis of William Herschel:

\textit{``Among the hypotheses that have been proposed on the origin of comets, the most probable seems to me to be that of Mr. Herschel, which consists in looking at them as small nebulae formed by the condensation of the nebulous matter diffused with such profusion in the universe Comets."}

Laplace believed that comets formed via the condensation of nebular material throughout the Galaxy and  subsequently traveled from stellar system to system. He calculated that 1/5713 comets within a 100000 au sphere around the Sun would exhibit perihelia $<2$ au with strongly hyperbolic trajectories. He concluded that the majority of interstellar comets would reach the Earth on parabolic orbits, consistent with the distribution of cometary orbits observed at that time. Unfortunately, Laplace neglected to include the proper motion of the Sun with respect to the galactic motion in his calculations.   Herschel had speculated that the sun moved with respect to  other stars in the Milky Way after observing the motion of several nearby stars decades previously \citep{Herschel1783}. 

It was not until 1866 and 1867 that Giovanni Schiaparelli redid the calculations from Laplace while including the proper Solar motion. He  demonstrated that the trajectories of interstellar comets would almost entirely be significantly hyperbolic. Therefore, while provocative in nature, the claim that interstellar comets were regularly seen throughout the Solar System was largely disregarded.

\subsection{Long and Short Period Comets}

 The comets in the Solar System are broadly categorized into two families: the Long Period Comets (LPCs) and the Short Period Comets (SPCs), the latter of which are commonly referred to as ecliptic comets or Jupiter Family Comets (JFCs). The LPCs arrive uniformly across the sky with an isotropic distribution of inclinations. This was attributed to   an isotropic cloud of comets as the source region of the LPCs, referred to as the Oort cloud \citep{Oort1950}. The Oort cloud spans 50000 au and, based on the occurrence of LPCs, it has been estimated to contain 10$^{11}$-10$^{12}$  km-sized Oort cloud objects \citep{francis2005,Brasser13,Dones15}. The total mass of the Oort cloud is typically quoted as 1 Earth mass (M$_\oplus$), but up to 20 M$_\oplus$ is allowable by the current data \citep{Kaib09}. 

Meanwhile, the tendency for the SPCs --- with orbital periods $<200$ yrs --- to lie within the ecliptic plane with low typical inclinations implies that these objects come from a different source region than the Oort cloud \citep{Everhart1972,Vaghi1973,Joss1973,Delsemme1973,Prialnik2020}.  It was therefore hypothesized that there existed a reservoir of icy objects past Neptune that sourced the SPCs. A small fraction of these trans-Neptunian objects would be gravitationally perturbed onto trajectories interior to the orbit  of Neptune. Subsequent dynamical interactions with the giant planets would transfer some of these objects into the inner Solar System\citep{Leonard1930,Edgeworth1943,Edgeworth1949,Kuiper1951,Cameron1962,Whipple1964,Fernandex1980MNRAS,Duncan1988,Quinn1990}. This trans-Neptunian population was verified in 1993 with the discovery of the first Kuiper belt object (KBO) (other than Pluto) \citep{Jewitt1993}.  Thousands of trans-Neptunian objects have since been identified and characterized \citep{Elliot2005,Volk2016,Shankman2016,Shankman2017,Bannister2018OSSOS}. It is now widely accepted that the JFCs are transported into the inner Solar System via the Centaur region between the giant planets \citep{Hahn1990,Levison1997,Tiscareno2003,DiSisto2007,Bailey2009,Sarid2019,Steckloff2020,Disisto2020}.

\subsection{Ejection of Interstellar Comets by the Solar System}

The existence of the Kuiper belt and Oort cloud has led to the realization that the Solar System underwent a period of planetary migration.  The migration of the giant planets specifically would have scattered a large amount of debris to large heliocentric distances, thereby populating the Kuiper belt and Oort cloud. There are varying hypotheses regarding the details of the planetary migration in the early Solar System. For a recent review of different formation models we refer the reader to \citep{Nesvorny2018}.  

Some authors have claimed that the Solar System underwent a transient period of violent dynamical instability in a model which is colloquially referred to as the ``Nice model" \citep{Tsiganis2005,Morbidelli2005}. Numerical experiments of this instability have demonstrated that it would have generated $\sim 30M_\oplus$ of material in interstellar comets. More general giant planet driven migration could also eject a similar amount of material \citep{Hahn1999,Gomes2004}. 

The majority of this material would be ejected in the form of interstellar comets. However, a fraction that has been estimated to be between 1-10$\%$ remained in the Solar System due to the effect of galactic tides and stellar flybys. These objects, now  protected from further interactions with the giant planets and from gravitational ejection, remain bound with eccentricity e $<1$ \citep{Hahn1999,Brasser10,Dones15,Higuchi15}. These processes likely generated the present day Kuiper belt and Oort cloud.

\begin{figure}%[h]
\begin{center}
       \includegraphics[scale=0.42,angle=0]{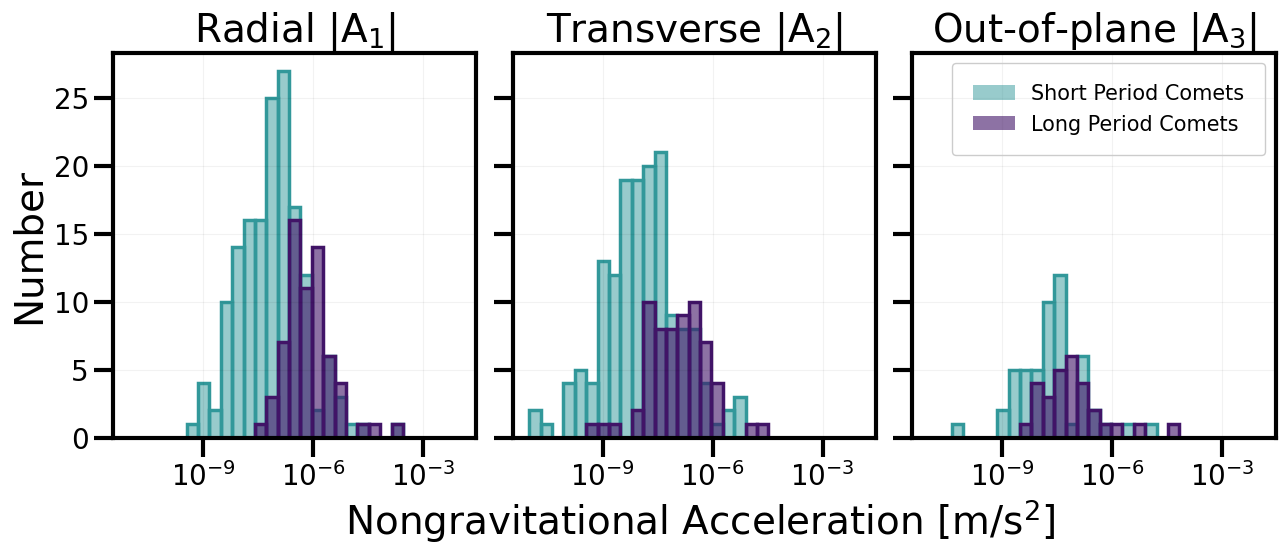}
    \caption{The measured nongravitational accelerations of Short-Period Comets (teal) and Long-Period Comets (purple). The three panels show the radial (left), transverse (middle) and out-of-plane (right) components of measured accelerations. Nongravitational accelerations  are taken from the JPL Small Body Database.}\label{fig:nongravs} 
\end{center}
\end{figure}

\subsection{Cometary Outgassing}\label{subsec:comets_outgassing}

Comets within the Solar System are generally detected via their distinct and explosive coma --- the gas surrounding the nucelus --- along with dusty cometary tails. These coma form when solar radiation received at the surface of the comet  produces an outflow of gas and dust. The energy from the solar irradiation  first powers the phase transition of the surface or subsurface ice into gas in a process called sublimation.  Next, the radiation energy is converted into kinetic energy which heats the gas to the thermal speed to power the outflowing jet. Dust particles and debris that are either residing on the surface of the comet or trapped within the ice typically travel along with the gas in the outflow. This entire process produces a cometary outflow and the dusty tail that, in some cases, is detectable from Earth.  Solar photons are then reflected off of the surface of these dust grains, which gives rise to the beautiful cometary tails. The wavelengths of light that are reflected depends on the size of the dust grains. For example, optical photons are reflected by $\sim$ micron-sized dust grains. This dust production can make the comet orders of magnitude brighter.

This mass loss produces a recoil effect on cometary nuclei \citep{Whipple1950,Whipple1951}. The force of the outflowing material can produce a nongravitational acceleration of the nucleus, typically written as $\vec{\alpha}_{ng}$.  A formalism to describe the nongravitational accelerations of comets was introduced by \citep{marsden1973} as,
\begin{equation}\label{eq:forces}
\vec{\alpha}_{\rm ng} = \bigg(\underbrace{ A_1 \hat{\mathbf r}}_\textrm{Radial} + \underbrace{A_2 \hat{\mathbf t}}_\textrm{Transverse} + \underbrace{A_3 \hat{\mathbf n}}_\textrm{Out-of-plane}\bigg) \, \underbrace{g(r_{H})}_\textrm{Radial dependance}\,.
\end{equation}
In Equation \ref{eq:forces},  $\hat{\mathbf r}$, $\hat{\mathbf t}$, $\hat{\mathbf n}$ are the  radial, transverse, and out-of-plane directions with respect to the orbit of the comet. The $g(r_H)$ function, or Marsden function,  is a function that parametrizes the dependence on the heliocentric distance $r_H$. The magnitudes of the three  components of the acceleration are  $A_1$, $A_2$, and $A_3$, normalized to  a heliocentric distance of $r_H=1$ au. 

This recoil acceleration is  related to the amount of mass lost in the outflow, written as $d M /dt$ with speed  $V_{s}$.  If the outflow is perfectly collimated in one direction --- typically thought to be normal to the surface ---  the outflow imparts momentum in one direction. If the outflow is completely isotropic then   the change of momentum perfectly cancels. We parametrize the degree of collimation in the outflow with the parameter $k_R$, where
 $k_R$ = 1 and $k_R$ =0 correspond to the  collimated and isotropic cases, respectively. Realistically, cometary outflows are somewhere between these two extremes and $0<k_R<1$. Assuming that the comet has a spherical nucleus with radius $r_n$ and  bulk density $\rho_\textrm{bulk}$, the mass loss can be written as,
 
 \begin{equation}
     \frac{dM} {dt} = \,\bigg(\,\underbrace{4\pi r_n^3 \rho_\textrm{bulk} \bigg/3}_\textrm{ Mass of Nucleus}\,\bigg)\,\bigg(\,\underbrace{ |\vec{\alpha}_{ng}|\bigg/k_R V_s}_\textrm{ Rate}\,\bigg)\,.
     \label{emdot}
 \end{equation}
  
The velocity of the outgassing species is related to the thermal speed. For example, H$_2$O ice interior to 2 au has a  sublimation temperature $T \sim$ 200 K which corresponds to a velocity $V_s \sim$ 500 m s$^{-1}$. This effect has been measured on a subset of comets in the Solar System. In Figure \ref{fig:nongravs}, we show measured nongravitational accelerations of Short and Long Period Comets.

\subsection{Composition of Comets}\label{subsec:comets_compositions}

Cometary tails are typically observed in optical wavelengths via the reflection of solar photons off of micron-sized dust particles in the outflows. However, emission and absorption of photons through the gas coma can also be measured spectroscopically, providing a direct probe of the \textit{volatile} content of the coma. Therefore, the composition of the sublimating gas itself in the coma has been directly measured in many cases. This volatile composition can be related to the bulk composition of the comet, with the caveat that it only represents the species active at the time of the observation. Subsurface or inactive surface ices of other species may be present in the nucleus but not the coma, thereby evading detection at any given point during an apparition.

\begin{figure}%[h]
\begin{center}
       \includegraphics[scale=0.35,angle=0]{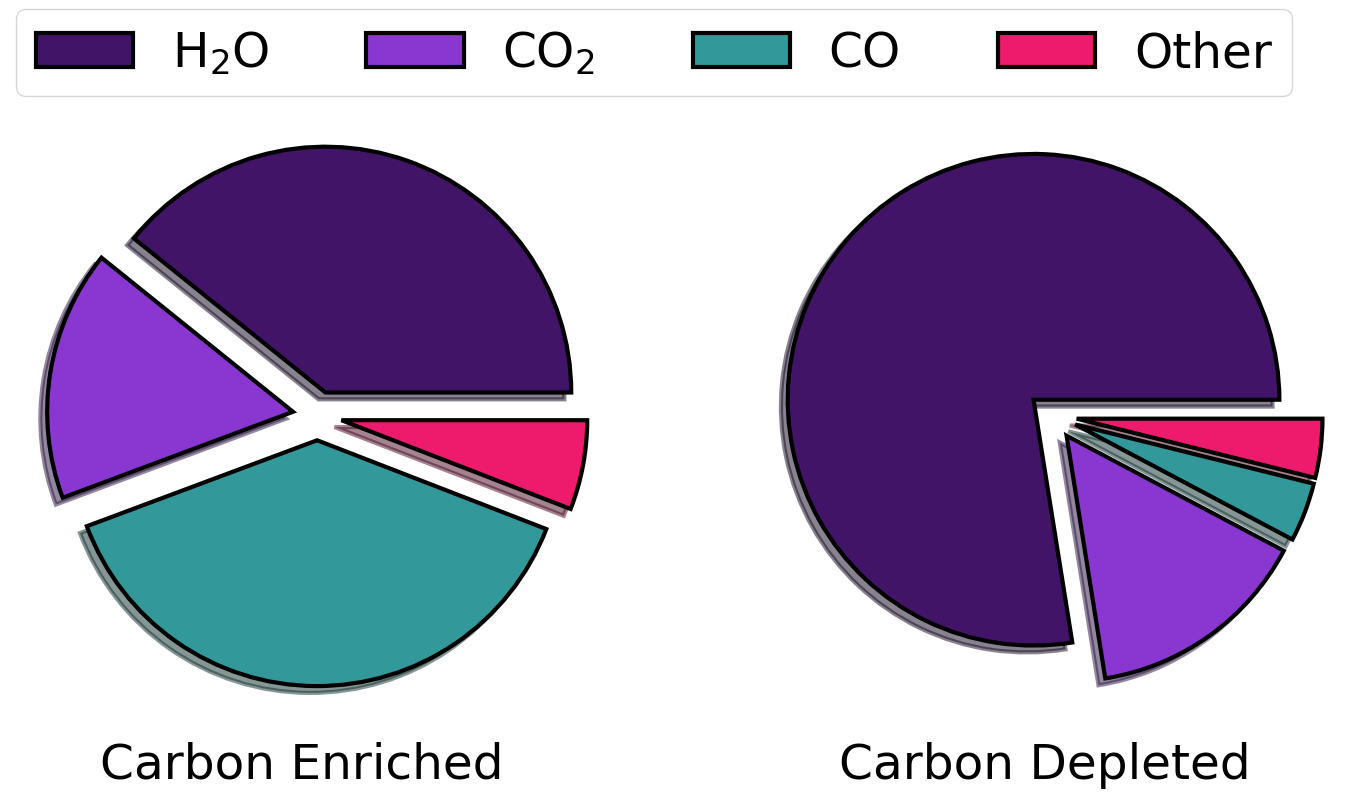}
    \caption{The typical compositions of carbon enriched and  depleted Solar System comets. This is a generalized version of an analogous figure in \citep{McKay2019}, and adapted from \citep{Seligman2022PSJ}. 
   The carbon depleted comet is representative of many of the Solar System comets for which production rate measurements of CO$_2$, CO and H$_2$O exist (see Table 1 in \citep{Seligman2022PSJ}).  The carbon enriched comet is C/2006 W3 Christensen \citep{Ootsubo2012}.  }\label{fig:Piecharts} 
\end{center}
\end{figure}

Despite these complications, the volatile content of a large set of comets has been measured. The majority of comets consist primarily of H$_2$O  ice \citep{Rickman2010,Cochran2015,Biver2016,Bockelee2017,Harrington22}.  The next two most abundant volatiles in comets are CO$_2$ and CO \citep{Harrington22}. Other carbon and nitrogen species can be abundant as well including CH$_4$, C$_2$H$_2$, C$_2$H$_6$, CH$_3$OH, NH$_3$ and HCN (see Figure 2 in \citep{Bockelee22}). In Figure \ref{fig:Piecharts}, we show typical compositions of carbon enriched and carbon depleted Solar System comets.

The composition of a comet can be linked to its formation location within the protoplanetary disk that initially formed from the Solar nebulae. The temperature of the ambient gas in the disk decreases with increasing distance from the Sun. Therefore, certain volatile species only freeze onto dust grains exterior to a characteristic temperature front, or snowline. For example, it is believed that in the protosolar disk, H$_2$O and CO were frozen exterior to the current day orbits of Jupiter and Neptune, respectively. An overview of cometary composition measurements, including measured production rates of CO, CO$_2$ and H$_2$O, was provided by \citep{Ahearn2012}. Based on the lack of hypervolatiles detected in the coma of most comets, they argued that the majority of comets formed in the giant planet region, where these hypervolatiles would not exist as ices to be incorporated into comets. The \textit{Akari} spacecraft provided a homogeneous survey of CO$_2$, CO and H$_2$O production rates in Solar System comets \citep{Ootsubo2012}. This provides a further line of evidence that the Kuiper belt and Oort cloud objects formed at closer heliocentric distances and were subsequently scattered outwards. However, it is important to note that this simplified picture does not incorporate the, potentially  significant, post-formation processing that also removes volatiles and hypervolatiles from comets.

    \section{1I/`Oumuamua}
    \subsection{Discovery}

The first interstellar object  was discovered by  Robert Weryk on 2017 October 19 from Haleakalā on Maui, Hawaii \citep{Williams17}. It was identified during the course of typical operations of the Pan-STARRs project \citep{Chambers2016}, a NASA funded all-sky survey optimized for detecting small Solar System bodies. The survey uses  two 1.8-meter telescopes and associated gigapixel detectors on the summit of Haleakalā. It is sensitive to a limiting magnitude of V = 22 and produces nightly images that have  $ \sim6000\,{\rm deg}^{2}$ resolution.

The object was initially flagged as a potentially hazardous near-Earth object (NEO) --- a population of small bodies from the asteroid belt that have been gravitationally perturbed onto Earth crossing trajectories. After only a few days of intensive and globally coordinated follow-up observations, the orbit of the object was confirmed to be hyperbolic with an eccentricity $e = 1.198$. The possibility that it was a Solar System object that had been gravitationally perturbed onto an unbound trajectory was quickly ruled out \citep{Wright2017,Schneider2017}. The hyperbolic  trajectory therefore confirmed that the object could not have originated from within our Solar System and definitively formed elsewhere. It  was discovered when it was $\sim$ 63 lunar distances from the Earth, after it had reached its closest approach to the Sun on 2017 September 9 and as it was rapidly departing the Solar System. The object had a closest approach distance to the Sun of $q\sim 0.255$ au and inclination $i = 122.8^\circ$.

There was explosive interest in the object after it was announced to be interstellar, both from the scientific community and from the general public.  Due to its close proximity to the Earth and its resulting rapid sky motion, the object was only observable for $\sim4$  weeks with ground based telescopes.  There was a global scramble to observe the rapidly moving object and director discretionary time was awarded to observe it on many major telescopes. The last successful ground based observation of the object was on 2017 November 21.  In the following subsection, we review all of the measurements obtained of the rapidly fading object.  For recent reviews, see \cite{Jewitt2022ARAA} and \cite{MoroMartin2022_review}.

    \subsection{Observations}\label{subsec:1I_observations}

    \begin{figure}
\begin{center}
       \includegraphics[scale=0.5,angle=0]{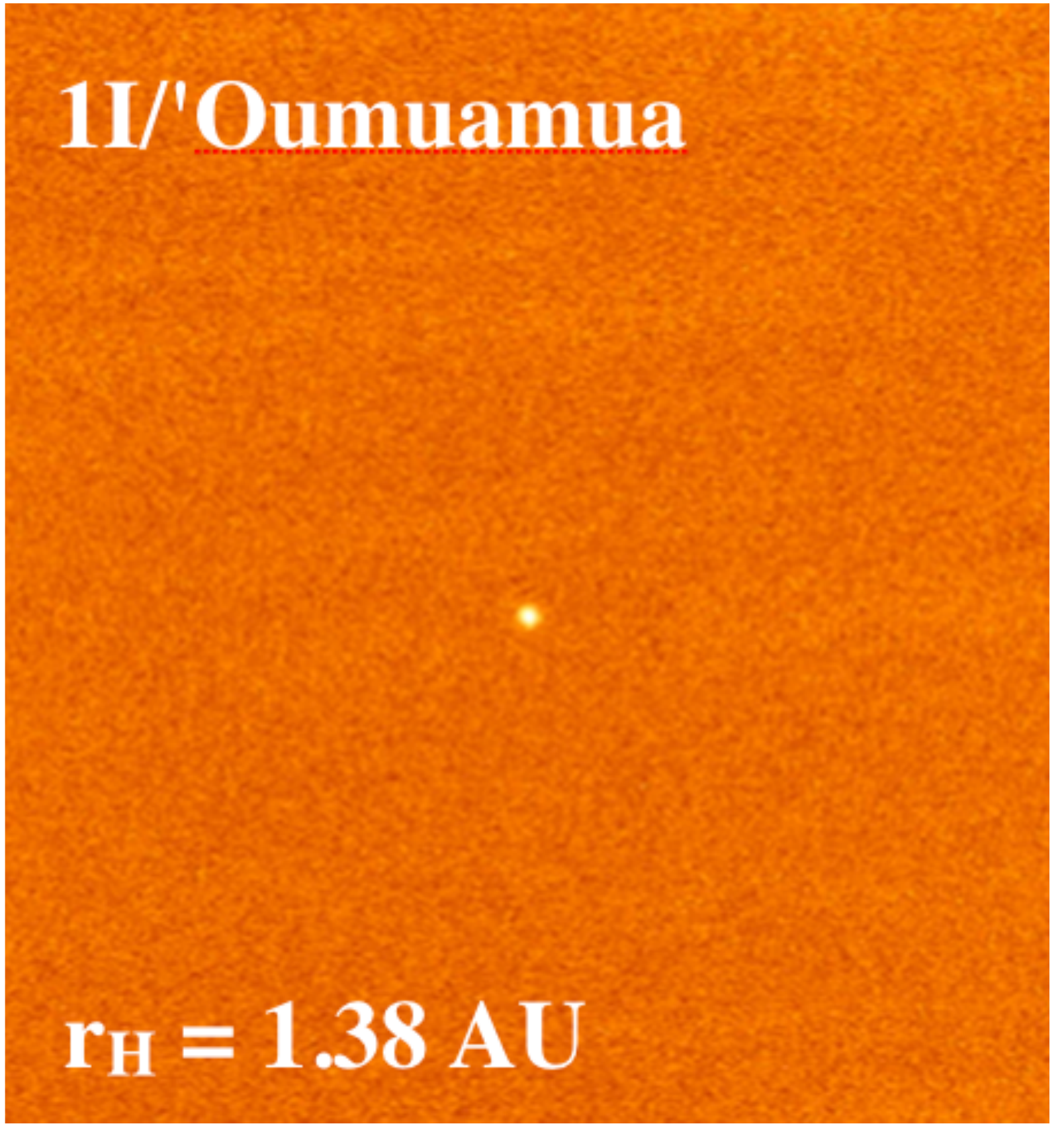}
    \caption{ An image of 1I/`Oumuamua taken with the 2.5 m Nordic Optical Telescope   on  2017 October 26. The object displays no distinct cometary tail.  In the figure, $r_H$ corresponds to the heliocentric distance of the object when the image was taken. This figure has been reproduced from \citep{Jewitt2017}.}
 \label{1I} 
\end{center}
\end{figure}

Immediately following the announcement of the discovery of the object, \citep{Masiero2017} obtained the first optical spectrum reflected from its surface on 2017 October 25 with the Palomar 5-meter Hale telescope.  The  spectrum spanned from 520 to 950 nm and was entirely featureless with a reddened slope. Further imaging and spectroscopic observations were taken of 1I/`Oumuamua the following night with the same telescope that indicated a consistently reddened surface color \citep{Ye2017}.

In a paper often credited with the discovery of 1I/`Oumuamua, \cite{Meech2017} reported time resolved photometric observations of the object from the nights of 2017 October 25, 26 and 27. These observations included 3.5 hr on the Very Large Telescope (VLT) 8-meter telescope,  3.5 hr on  Gemini south, 3 hr on the Keck 10-meter telescope, 9 hr on the United Kingdom Infrared Telescope (UKIRT),   and 8 hr on the Canada–France–Hawaii telescope (CFHT). Deep stacks of the images showed no evidence of cometary activity or dust tail.

In \citep{Jewitt2017}, observations were reported on the nights of 2017 October 25, 26, 29 and 30 with the  2.5-meter diameter Nordic Optical Telescope (NOT) and 4.5 hr on the WIYN telescope, located at Kitt Peak National Observatory in Arizona. They measured the optical colors of the nucleus to be B-V $= 0.70\pm0.06$, V-R $= 0.45\pm0.05$ and found a lack of micron-sized dust particles.  Figure \ref{1I} shows an image of 1I/`Oumuamua taken with the NOT telescope, clearly lacking a dust tail and apparent coma.

Spectroscopic observations of 1I/`Oumuamua were reported in \citep{Fitzsimmons2017}, with data from the 4.2-meter William Herschel Telescope (WHT) on La Palma on 2017 October 25 and the X-shooter spectrograph located at the European Southern Observatory 8.2 m Very Large Telescope (VLT) on 2017 October 27. The resulting spectrum covered 0.3–2.5 $\mu$m, and was also featureless and reddened. Photometric observations (8.06 hr) were reported by \citep{Drahus18} obtained with the Gemini North 8.1-meter telescope on 2017 October 27 and 28 with extremely high temporal cadence, resulting in 431 r'-band images. Additional spectroscopic observations were described by \citep{Bannister2017} obtained  with the Gemini telescope (2 hr) and the 4.2 m William Herschel Telescope (WHT) (2 hr), on 2017 October 29 and 30, respectively.  One those same dates, \citep{Bolin2017} obtained observations (4 hr)  using the  Apache Point Observatory (APO) ARCTIC camera on the 3.5-meter telescope  (2017 October 29) and \citep{Knight2017} obtained photometric data (2.8 hr) with the Lowell Observatory 4.3-meter Discovery Channel Telescope (DDT) (2017 October 30). 

The NASA/ESA \textit{Hubble Space Telescope} (HST) alloted 9 orbits of  time  to monitor 1I/`Oumuamua. These data were taken with the UVIS channel from the Wide-Field Camera 3 (WFC3) using the F350LP filter.  These observations occurred during 2017 November, December and 2018 January over all 9 orbits. Photometric brightness variations were reported  during 2017 November 21 and 22, and the remaining observations provided astrometric measurements as well \citep{Belton2018}. They  also presented data in the same paper with the CFHT MegaCam imager from 2017 November 22 and 23 and  the Magellan-Baade 6.5-meter telescope at Las Campanas Observatory from 2017 November 21, 22 and 23. They compiled a composite light curve over a 29.3 day period spanning about 0.13 au of the trajectory  containing a total of 818 photometric observations of the object reported by \citep{Meech2017,Bolin2017,Bannister2017,Drahus18,Fraser2017,Jewitt2017,Knight2017,Belton2018}.

\begin{table}[t]
\begin{center}
\caption{Summary of all measured upper limits on  production rates of various volatile species and dust from 1I/`Oumuamua. Note that the measurement indicated with $^*$ had a typographical error which was later fixed. }\label{table:production}
\begin{tabular}{ ccccc} 
Species  & Physical Property & Value& Distance& Reference  \\
\hline
&&&&\\
{\rm CN } & Q({\rm CN})[molec s$^{-1}$ ]& $<2\times10^{22}$&1.4 au &\citep{Ye2017}\\
{\rm C$_2$ } & Q({\rm C$_2$})[molec s$^{-1}$ ]& $<4\times10^{22}$&1.4 au &\citep{Ye2017}\\
{\rm C$_3$ } & Q({\rm C$_3$})[molec s$^{-1}$ ]& $<2\times10^{21}$&1.4 au &\citep{Ye2017}\\
{\rm OH } & Q({\rm OH})[molec s$^{-1}$ ]& $<1.7\times10^{27}$&1.8 au &\citep{Park2018}\\
{\rm CO$_2$ } & Q({\rm CO$_2$})[molec s$^{-1}$ ]& $<9\times10^{22}$&2.0 au &\citep{Trilling2018}\\
{\rm CO$^*$ } & Q({\rm CO})[molec s$^{-1}$ ]& $<9\times10^{21}$&2.0 au &\citep{Trilling2018}\\
{\rm CO } & Q({\rm CO}) [molec s$^{-1}$ ]& $<9\times10^{23}$&2.0 au &\citep{Seligman2021}\\
Dust  & Q({\rm Dust})[kg s$^{-1}$ ]& $<1.7\times10^{-3}$&1.4 au &\citep{Meech2017}\\
Dust  & Q({\rm Dust})[kg s$^{-1}$ ]& $<2\times10^{-4}$&1.4 au &\citep{Jewitt2017}\\
Dust  & Q({\rm Dust})[kg s$^{-1}$ ]& $<10$&1.4 au &\citep{Ye2017}\\
\hline
\end{tabular}
\end{center}
\end{table}

In addition to these detections of 1I/`Oumuamua, several attempts to observe the object resulted in non-detections. Some of these provided critical upper limits on the  production rates of dust and certain volatile species. \citep{Ye2017} attempted to observe 1I/`Oumuamua with the Canadian Meteor Orbit Radar, and the non-detection provided upper limits on the production of  CN, C$_2$ and C$_3$ (Table \ref{table:production}). An upper limit on the production rate of OH via the OH 18-cm line (Table \ref{table:production}) were reported by \citep{Park2018} based on 4 hr of observations with the Green Bank Telescope on 2017 November 12.

The \textit{Spitzer Space Telescope} observed 1I/`Oumuamua for 32.6 hr on 2017 November 21 and 22. These observations were taken with the 4.5 $\mu$m band in the IRAC channel 2, which would have detected fluoresence of carbon-based molecules such as CO or CO$_2$. However, these observations resulted in a non-detection which provided an upper limit on the production of CO and CO$_2$ of the object \citep{Trilling2018} (Table \ref{table:production}). Note that the production of CO contained a typographical error and was later corrected by \citep{Seligman2021}. 

A precovery of 1I/`Oumuamau prior to perihelion from 2017 September was attempted by \citep{Hui2019}. Their search for the object in extant data from Solar TErrestrial RElations Observatory (STEREO) and the Solar and Heliospheric Observatory (SOHO) resulted in nondetections. 

Several teams attempted to detect radio signal from 1I/`Oumuamua, all of which were unsuccessful. Data were obtained with the  Robert C. Byrd Green Bank Telescope beginning on  2017 December 13  \citep{Enriquez2018}, the  Murchison Widefield Array (MWA) on 2017 November 28  \citep{Tingay2018}  and the Allen Telescope Array from 2017 November 23 through December 5 \citep{Harp2019}. All of these upper limits on production rates are summarized in Table \ref{table:production}.

\subsection{Nucleus Properties}

\subsubsection{Size}

The size of asteroids can be related to their absolute magnitude \citep{Pravec2007} with the following relationship,

\begin{equation}\label{eq:diameter_mag}
    2 \bigg(\,\frac{r_{\rm n}}{1 \textrm{km}}\,\bigg) =\,\bigg(\, \frac{1329}{\sqrt{p}} \,\bigg)\, 10^{-0.2 \, H}\,.
\end{equation}
In Equation \ref{eq:diameter_mag}, $H$ is the absolute magnitude inferred from the brightness and geometry of the orbit (H = 22.4 for 1I/`Oumuamua), $p$ is the geometric albedo and $r_{\rm n}$ is the radius in kilometers of the assumed to be spherical nucleus. For the case of asymmetrical shaped objects, $r_{\rm n}$ is estimated as the effective radius. This is approximated as the circular cross-section of an ellipse with aspect ratio $a:b$,  and therefore $r_{\rm n} = (ab)^{1/2}$.  The albedo is a  measure of the fraction of light reflected off of the surface versus absorbed.  The albedo of nitrogen ice seen on the surface of Pluto can be as high as $p=0.8-0.9$, while H$_2$O ice on the surface of Charon has a more moderate albedo $p\sim0.25$ \citep{Buratti2017}. However, the albedos of carbonaceous surfaces of  asteroids are much smaller with $p \sim$ 0.04.  

The surface composition of 1I/`Oumuamua is unconstrained and therefore the exact size of the object is not known. Literature estimates of the size varied from 55 m \citep{Jewitt2017}, 70$\pm$3 m \citep{Meech2017}, 80 m \citep{Drahus18} to 114 m \citep{Knight2017}. Assuming an albedo $p=0.1$, \citep{Jewitt2022ARAA} estimated an effective radius of 80 m. This value is consistent with the nuclear radius calculated with Equation \ref{eq:diameter_mag} using an absolute magnitude of $H = 22.4$.

\subsubsection{Colors}\label{subsec:colors}

The surface colors of small bodies can be measured spectroscopically. These colors do not give definitive information of the surface compositions but can be used to identify  surface features.  For example, asteroids in the inner Solar System have reddish colors that are due to iron produced by particle bombardment.   

Most objects in the Solar System have slightly reddened reflectance spectra, with the exception of cold classical KBOs and some Centaurs. The inclination distribution of the classical KBOs is known to be bimodal. Objects with low inclinations $i<5^\circ$ are referred to as the cold classical KBOs, while objects with larger inclinations are hot classical KBOs \citep{Brown2001,Gulbis2010}. The cold classical KBOs likely formed in situ and therefore have  little or no residency time in the inner Solar System \citep{Nesvorny2018}. These cold classical KBOs and some inactive Centaurs have surface reflectance spectra consistent with ``ultra-red" material \citep{Hainaut2012,Jewitt2015,Jewitt2018}. It is believed that the ultra-red material is caused by carbon compounds that are unstable and depleted in the warmer inner Solar System, and produced via space weathering from cosmic rays and interstellar medium (ISM) plasma \citep{Jewitt2017}. 
 
 Therefore, a natural hypothesis is that the surface colors of interstellar objects that are exposed to harsh galactic environments  should resemble the surfaces of cold classical KBOs. Somewhat surprisingly, published measurements of the color of 1I/`Oumuamua are more broadly consistent with the material in the inner Solar System and the object lacked ultra-red material. 

    \subsubsection{Shape}
For the majority of Solar System objects, photometric light curves provide the only information available regarding the shape of the object. Sometimes the shape models inferred from light curves can be verified and/or updated with radar measurements. Notable examples include the ``dog-bone'' shaped asteroid (216) Kleopatra \citep{Ostro2000} and 1998 KY$_{26}$ \citep{Ostro1999}. In  special cases, in-situ space based measurements reveal detailed geometry and terrain features. Therefore, much effort has gone into interpreting the shape of an object from photometric light curves alone.

    The composite light curve of 1I/`Oumuamua displays several salient features \citep{Belton2018,Jewitt2022ARAA}. The first is that the brightness varied  by a factor of about $\sim12$ --- note that the y-axis in most published depictions of the lightcurve figure is in the scale of visual magnitudes, which is a logarithmic scale. At the time of the detection of 1I/`Oumuamua, no object in the Solar System had displayed such extreme amplitude brightness variations. It is worth noting that since the discovery of 1I/`Oumuamua,  the NEO 2016 AK$_{193}$ exhibited brightness variations of $\sim2.5-3$ magnitudes during its discovery apparition and in subsequent followup observations \citep{Heinze2021DPS}. Measurements of 2021 NY$_{1}$ and 2022 AB were reported by \citep{Licandro2023}  indicating aspect ratios of $\ge 3.6$ and $\ge1.6$ respectively.
    
    The effects of both the orbital configuration and the geometry of an object can lead to amplitude variations, which can be difficult to disentangle. The phase angle, or the angle between the object sun and observer, can increase the amplitude of the light curve variations. Lightcurve variations are intrinsically driven by two features: an (i) elongated shape of the body and/or a (ii) nonuniform reflectance (or abledo) across the surface of the object. In the case of 1I/`Oumuamua, the aspect ratio implied by the lightcurve is $6:1$ when correcting for the phase of the object throughout its trajectory. However, literature estimated of the aspect ratio varied from $5:3:1$ \citep{Bannister2017}, $6:1$ by \citep{Jewitt2017,McNeill2018} to $10:1$ \citep{Meech2017}. It was unclear whether the elongation corresponded to a prolate or oblate ellipsodal geometry at the time of the discovery. A rigorous and exhaustive fit to the light curve using $\sim1$ GPU-year of computational time was performed by \citep{Mashchenko2019} demonstrating that the oblate or ``thin-disk" geometry with a $6:1$ aspect ratio was favored with a $91\%$ likelihood. In retrospect, this geometry is intuitive because only an oblate geometry would produce deep minima during each circulation. The extreme shape was suggested to be the result of continuous bombardment and abrasion by interstellar dust particles \citep{Domokos2009,Domokos2017,Vavilov2019}, a violent collision \citep{Sugiura2019}, a tidal disruption of a larger object that came too close to a giant planet \citep{Raymond18} or its parent star \citep{Zhang20} before it was ejected to interstellar space, or a fractal aggregate structure \citep{moro2019fractal}. It should be noted that a geometry of an ultra-thin minor axis was shown to produce a low ($\sim$1\%) probability of generating the observed large amplitude \citep{Zhou2022}.

    To recreate these types of variations from only albedo features, the object would need to have one hemisphere darkened with respect to the other. Space weathering and processing in the ISM  would have been approximately isotropic. Therefore, this interpretation has been  more or less disregarded. However, alternative reflection models such as Lambertian or specular reflection and single scattering diffusive and backscatter could plausibly reproduce the observed brightness variations with a less extreme aspect ratio \citep{Vazan2020}.
    
    The second salient feature is that there is an approximately constant periodicity of $\sim8$ hr. If the object was elongated and rotating then each cycle would correspond to complete circulations as the objected reflected sunlight from the maximum and minimum cross-sectional areas.  While the amplitude variations maintain constant periodicity, each individual cycle exhibits significant asymmetries. This was highlighted in \cite{Drahus18}, who showed phase-folded photometric data obtained with Gemini North telescope on 2017 October 27 and 28. The folded data show $\sim$10\% differences at each circulation. This was interpreted as the object existing in an excited or tumbling rotational state \cite{Fraser2017,Drahus18,Belton2018}

    \subsection{Nongravitational Acceleration}

Despite the lack of visible dust coma or detected outgassing species, \cite{Micheli2018} reported a 30-$\sigma$ significant detection of nongravitational acceleration in 1I/`Oumuamua's trajectory. This  acceleration was based on astrometric fitting over a 2.5 month arc of the orbit.

The only significant component of the nongravitational acceleration of 1I/`Oumuamua was the radial component A$_1$, where $A_{1}\sim2.5\times10^{-4}\,{\rm cm\,s^{-2}}$ at $r\sim1.4\,~{\rm au}$ to a 30-$\sigma$ significance. The best fitting  $g(r_H)$ functions for the nongravitational acceleration were those that depended inversely on the distance with exponents ranging from $r_H^{-1}$ to $r_H^{-2}$. It is important to note that the exponent could not be constrained any better than being somewhere in the -1 to -2 range. The magnitudes of the best fitting nongravitational acceleration components in the transverse and normal directions were consistent with zero. In \cite{Micheli2018} it was concluded that the most likely cause of the nongravitational acceleration was  outgassing. While some asteroids exhibit nongravitational accelerations caused by the Yarkovsky effect and radiation pressure (see subsections\ref{subsec:asteroids_active} and \ref{subsec:asteroids_darkcomets}), the magnitude of the acceleration was too large to be explained by these effects. The nongravitational acceleration combined with the lack of visible coma led to a wide array of hypotheses regarding the provenance of the object which we review in the following subsection.

\subsection{The Provenance of 1I/`Oumuamua}

\subsubsection{Radiation Pressure-Driven Fractal Aggregate or Membrane}

 The two main causes of nongravitational accelerations seen on asteroids are the Yarkovsky effect \citep{Vokrouhlicky2015_ast4} and solar radiation pressure \citep{Vokrouhlicky2000}. The Yarkovsky effect results from anisotropic reradiation of thermal photons and typically manifests in the transverse $\hat{\mathbf t}$ direction.  Solar radiation pressure is only detected in the  radial $\hat{\mathbf r}$ direction and is weaker than outgassing effects but stronger than Yarkovsky effects for typical size and compositions of asteroids. Therefore it has only been measured  on a handful of small asteroids \citep{Micheli2012,Micheli2013,Micheli2014,Mommert2014bd,Mommert2014md,Farnocchia2017TC25,Fedorets2020}. For a more detailed discussion, we refer the reader to Subsections \ref{subsec:asteroids_active} and \ref{subsec:asteroids_darkcomets}.

Radiation pressure is therefore a natural explanation for the source of the acceleration of 1I/`Oumuamua, which was only significant in the radial direction. However, this was originally considered and dismissed by \citep{Micheli2018} because it would imply an extremely low bulk density or minor axis thickness. Given the lack of detectable dust coma, the hypothesis of radiation pressure induced acceleration was reconsidered by several authors. The theories that were published invoking radiation pressure can be categorized by those invoking (i) a naturally produced, ultra-low density aggregate \citep{moro2019fractal,Sekanina2019b,Luu20} or (ii) a membrane-like structure of a very thin material (less than a mm thick) of an unknown natural or artificial origin, similar to a lightsail \citep{Bialy2018}. 

Regarding the former, a fractal aggregate with a density of $\sim 10^{-5}$ g cm$^{-3}$ (about 100 times less than air) could provide the surface-to-mass ratio required to account for the observed acceleration of 1I/`Oumuamua \citep{moro2019fractal,Sekanina2019b}. A fractal structure could also help explain its unusual shape. However, the lowest density solid known (10 times less dense than air) is graphene aerogel and is synthetically produced. Therefore, it is unclear how such an ultra-porous structure could be naturally produced. 

Fractal aggregates are found in many forms of nature and are thought to arise naturally because their formation processes involve an element of stochasticity. In a protostellar disk, the relative motions of neighboring dust particles are small and therefore collisions are gentle. These conditions may be amenable to diffusion-limited aggregation \citep{Witten1981}. Numerical simulations investigating planetesimal growth via dust collisions in a protoplanetary disk show that beyond  icelines, if the tiny dust particles are covered with ice and are about 0.1 $\mu$m in size, the aggregates that form will have increasingly smaller densities as they grow \citep{Suyama2008,Okuzumi2012}. This occurs  because in the early phases when the particles are small and  strongly coupled to the gas, the relative velocities of the colliding aggregates is  low. Therefore, the collisions   are not  energetic enough to restructure the aggregates. This leads to a rapid increase of  porosity as the aggregate grows. As the aggregates become larger, collisional compression occurs. However,  this compression is inefficient and the porosity of the growing aggregate continues to increase. This increase occurs because most of the colliding energy is spent compressing the new voids that are created when two aggregates collide and stick to each other, rather than compressing the voids that were already present in the colliding aggregates (see review by \citep{Kataoka2017}). What these studies show is that at $\sim$ 100 meters, about the size of 1I’/Oumuamua, the planetesimals can obtain  ultra-low densities that are comparable to what would be required for 1I/`Oumuamua to be “pushed” by radiation pressure \citep{Okuzumi2012}. It was therefore suggested that 1I/`Oumuamua could be one of these intermediate products of the planet formation process  \citep{moro2019fractal}.

If this were its origin, it would be extraordinarily exciting because very little is known about these intermediate/early products of planet formation. Studying these objects could help  set unprecedented constraints on planet formation models. In particular, the intermediate stage of planet formation, in which cm-sized particles grow into km-sized planetesimals, is not well understood because that process should theoretically  be very inefficient. Several mechanisms have been proposed to help overcome this ``meter-sized barrier” and one of them is  the presence of very high porosity planetesimals. The high porosity could favor the growth process because  the larger cross section \citep{Suyama2008}  would allow a longer lifetime against the effect of radial drift in the inner regions of the protoplanetary disk. This would both facilitate growth \citep{Okuzumi2012} and   make the colliding aggregates less susceptible to fragmentation and bouncing.

There are other origins that have been been proposed for an ultra-low density 1I/‘Oumuamua. Coma aggregates collected from 67P/Churyumov-Gerasimenko with the ESA Rosetta mission had low densities $<$1 kg m$^{-3}$ \citep{Fulle15} and fractal structure \citep{Mannel16}. It was therefore hypothesized that 1I/`Oumuamua was a similar but larger-scale fractal aggregate formed in the coma of an active comet, that then escaped \citep{Luu20}. Another suggestion is that it was an ultra-porous desicated fragment that resulted from the disintegration of an “ordinary” and larger interstellar comet as it passed near perihelion \citep{Sekanina2019b}.

Regarding the survival of such an ultra-porous structure to the hazards of interstellar travel, somewhat counter-intuitively, \citep{Flekkoy19} demonstrated that it could be stable to tidal disruption. Furthermore, they argued that the interaction of such an ultra-low density aggregate with the solar radiation could explain the changes observed in the rotational period of 1I/‘Oumuamua. It remains to be studied if this fractal structure could survive the ejection from its parent system and the passage near the Sun.

\subsubsection{Tidally-Disrupted Planetesimal or Planet Fragment}

The extreme aspect ratio of 1I/`Oumuamua remains somewhat of a mystery. In another line of reasoning, several authors  considered that its extreme shape was primordial and caused by tidal forces. Tidal forces can have significant effects on small bodies like comets and asteroids. These manifest as body deformation in the mildest cases and catastrophic disruption in the most extreme cases.  A noteable example is the tidal disruption of the comet Shoemaker-Levy 9 when it passed close to Jupiter \citep{Shoemaker1993, Weaver1995,Lellouch1995,Noll1995}.  Chains of craters  on the surfaces of Callisto and Ganymede may  have been caused by catastrophic disruption of objects   that subsequently impacted these surfaces \citep{Schenk1996}. Double craters that are seen on Solar System bodies \citep{Melosh1991,Bottke1996,Bottke1996b,Melosh1996,Cook2003} may have been caused by  objects that underwent satellite formation from tidal disruption \citep{Richardson1998}.

If 1I/`Oumuamua was ejected from its host system following a close encounter with a giant planet or its host star, it is possible that it too was subject to tidal forces that produced an elongated fragment. The dynamical simulations in \citep{Raymond18,Rayond2018MNRAS,Raymond2020} show that $\sim$1\% of planetesimals pass within the tidal disruption radius of a gas giant on their pathway to ejection. If 1I/`Oumuamua was ejected from its host system following such a close encounter with a giant planet, it is possible that it too was subject to tidal forces, producing an elongated fragment \citep{Raymond18,Rayond2018MNRAS,Raymond2020}. Furthermore, these studies argued that these fragments might have a characteristic size of $\sim$ 100 m, similar to 1I/`Oumuamua’s size, as opposed to a wider size distribution. Alternately, \cite{Zhang20} proposed that 1I/'Oumuamua is a fragment that resulted from the tidal disruption of a planet, or a small body, that came too close to its parent star and that was later ejected to interstellar space. Based on detailed numerical simulations for a range of impact parameter and material properties, they suggested this process could produce a prolate object with the observed aspect ratio. 

However, \citep{Mashchenko2019} demonstrated that the oblate shape was by far the most likely geometry of the object, which casts into doubt both tidal fragment hypothesis as the source of the extreme geometry. Moreover, \citep{Taylor2022} demonstrated that the lightcurve of 1I/`Oumuamua was consistent with no tidal deformation during its observed passage through the inner Solar System.

\subsubsection{Sublimating Icy Comet}

In \citep{Micheli2018}, it was argued that the most natural explanation for the nongravitational acceleration was cometary outgassing. However,  \citep{Rafikov2018} compared 1I/`Oumuamua to comets for which the rotational periods of the nuclei had been measured and argued that the magnitude of the nongravitaitonal acceleration should have caused a measurable change in the nuclear spin period.  Using an outgassing model, \citep{Seligman2019} argued that a rapidly mobilizing outflow that tracked the point of maximal solar irradiance would not result in significant spin-up of the object. However, this model is highly idealized and in reality cometary nuclei do experience secular spin-up that can lead to disintegration \citep{Jewitt2003b,Drahus2011,Gicquel2012,Maquet2012,Fernandez2013,Wilson2017,Eisner2017,Roth2018,Kokotanekova2018,Biver2019,Combi2020,Jewitt2021,Jewitt22}. In \citep{Flekkoy19}, it was reported that in the high-cadence data of 2017 October, 1I/`Oumuamua experienced a slight spin-down in its rotational period that could be attributed to the YORP effect on a fractal body with low ultra-low density. 

A further issue with the outgassing hypothesis is that 1I/`Oumuamua did not receieve sufficient energy in the form of solar irradiation to power typical outgassing of H$_2$O or CO$_2$ (see subsections \ref{subsec:comets_outgassing} and \ref{subsec:comets_compositions}) \citep{Sekanina2019}. The sublimation rate per unit surface area can be calculated as an equilibrium solution (as described in Section 4.1 of \citep{Jewitt2022ARAA}) to the equation

\begin{equation}
\bigg(\,\underbrace{\frac{ (1-p)\,L_{\odot}}{4\pi r_H^2}}_\textrm{Solar energy}\,\bigg)\, k_R =\, \underbrace{\varepsilon \sigma T_{S}^4}_\textrm{Re-radiated} + \underbrace{H f_s(T_S)}_\textrm{Sublimation} + \underbrace{C(T_S)}_\textrm{Interior}\,.
\label{equilibrium}
\end{equation}
In Equation \ref{equilibrium}, $p$ is the albedo,  $r_H$ the heliocentric distance, $L_{\odot}$ is the Solar luminosity, $\varepsilon$ is the surface thermal emissivity, $\sigma$  is the Stefan-Boltzmann constant, $T_{S}$ (K) is the  temperature at the surface, $H$  is the latent heat of sublimation, $f_s(T)$  is the  sublimation rate and $k_R$ is the degree of collimation of the outflow described in Equation \ref{emdot}. The LHS of Equation \ref{equilibrium} is the total energy received from solar radiation, and the RHS indicates the total energy required to produce the outflow. On the RHS, $\varepsilon \sigma T_{S}^4$ is the energy reradiated from the surface of the comet, $H f_s(T_S) $ is the energy deposited in the form of the latent heat of sublimation to convert the ice into a gas, and $C(T_S)$ represents heat conduction into the interior. The last term   is negligible in most cases.

Equation \ref{emdot} can be solved using both the nongravitational acceleration magnitude and fiducial size of 1I/`Oumuamua. This produces  a required sublimation rate of $\dot{M} =$ 24 kg s$^{-1}$ needed to provide the acceleration \citep{Jewitt2022ARAA}. From Equation \ref{equilibrium}, H$_2$O and CO$_2$ are  insufficient to provide this sublimation for 1I/`Oumuamua due to the relatively high latent of these species.  Since  H$_2$O and CO$_2$ are the most common volatiles seen in Solar System comets (subsection \ref{subsec:comets_compositions}), if 1I/`Oumuamua was outgassing it is unlike most icy objects we have seen in the Solar System. \citep{Seligman2020} demonstrated that when considering these bulk energetics, only hypervolatiles  such as molecular hydrogen H$_2$, molecular nitrogen N$_2$, CO or more exotic species such as Xenon, Neon, Krypton or Argon could have been feasible accelerants. The noble gases, however, exhibit relatively low cosmic abundance and are therefore a priori not likely to be bulk constituents of 1I/`Oumuamua.

\begin{figure} 
\begin{center}
       \includegraphics[scale=0.4,angle=0]{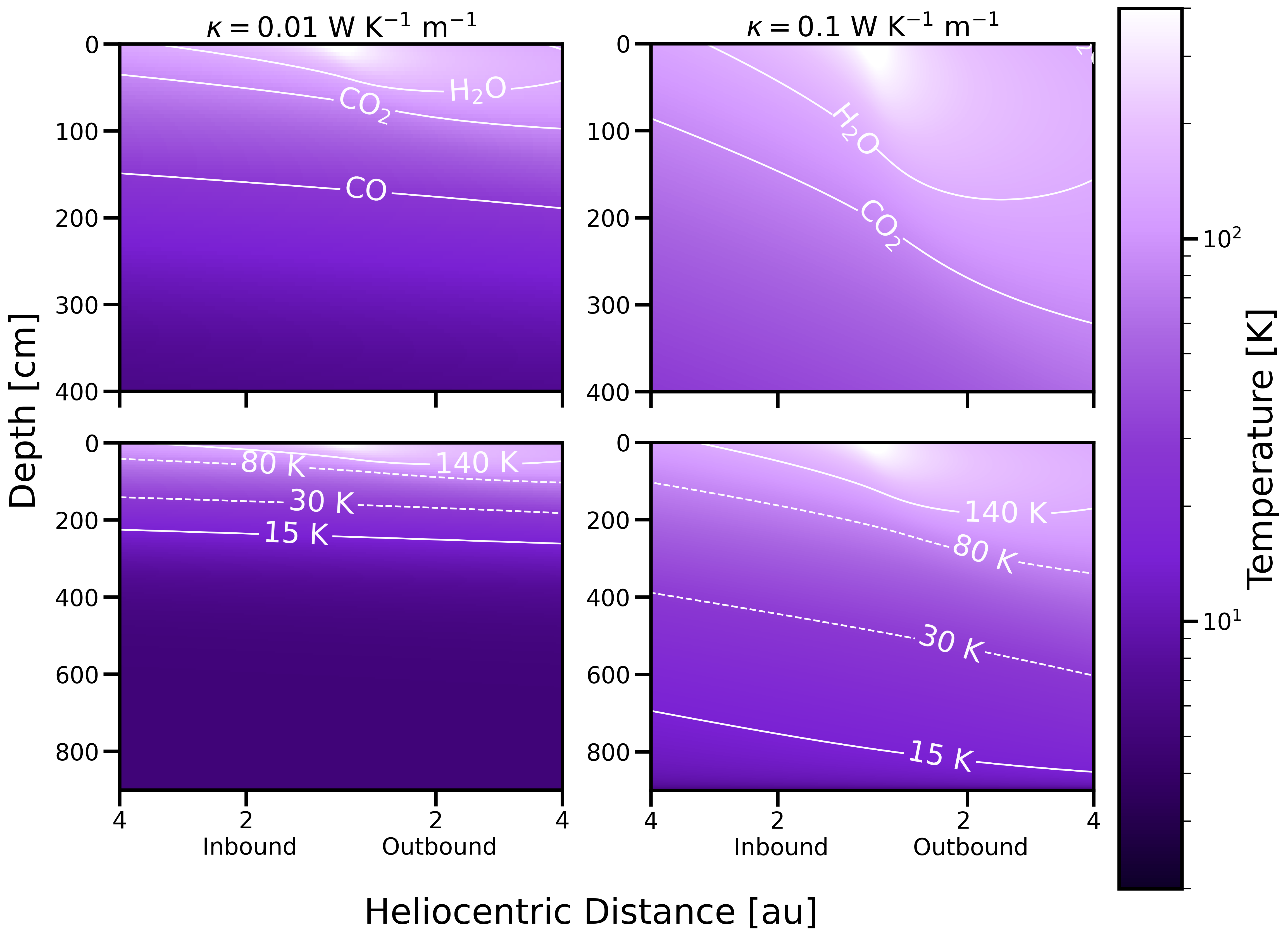}
    \caption{Numerical solutions of the radial heat equation (Equation \ref{eq:CylHeatEq}) for 1I/`Oumuamua spanning the inbound and outbound trajectory interior to 4 au.  We adopt a bulk density $\rho_{Bulk}=0.5$ g cm$^{-3}$, specific heat capacity of $c_P=2000$ J kg$^{-1}$ K and thermal conductivity $\kappa=10^{-2}$ W K$^{-1}$ m (left panels) and $\kappa=10^{-1}$ W K$^{-1}$ m (right panels).  Contours on the left panels indicate sublimation fronts of CO (28 K), CO$_2$ (86 K) and H$_2$O (144 K) \citep{Gasc2017}. Contours on the right panels indicate where H$_2$ outgassing is expected during the annealing of amorphous H$_2$O ice for temperatures of $\sim$15--140 K. Adapted from \citep{Bergner2023}. }\label{fig:thermal} 
\end{center}
\end{figure}

The subsurface layers of 1I/`Oumuamua are required to reach sufficient  temperatures for the outgassing volatiles to sublimate in order to power the  acceleration. In \citep{Jewitt2017}, it was first estimated that the thermal skin depth of 1I/`Oumuamua was $\sim0.5$ m via an order of magnitude calculation to argue that the internal temperature could remain at $\sim10$ K throughout the trajectory.  In \citep{Fitzsimmons2017} a more detailed numerical thermal model was presented that demonstrated that H$_2$O was stable to depths of $\gtrsim$30 cm (see Figure 4 in \citep{Fitzsimmons2017}). The thermal modelling of the interior requires numerical solutions to the heat conduction equation,
 \begin{equation}\label{eq:CylHeatEq}
    \frac{dT}{dt}=\frac{\kappa}{\rho_{Bulk} c_P}\frac{1}{z}\frac{\partial}{\partial z}\,\bigg(\,z\frac{\partial T}{\partial z}\,\bigg) \, .
\end{equation}
In Equation \ref{eq:CylHeatEq},  $T$ is the   temperature,  $\kappa$ is the thermal conductivity,  $c_P$ is the specific heat capacity, and $z$ is the  depth. The thermal properties of 1I/`Oumuamua are entirely unconstrained, but typical cometary properties are  $\rho_{Bulk}=0.5$ g cm$^{-3}$ \citep{britt2006}, $c_P=2000$ J kg$^{-1}$ and $\kappa=10^{-2}-10^{-1}$ W K$^{-1}$ m$^{-1}$ \citep{Steckloff2021}. We show the results of this thermal modelling for 1I/`Oumuamua in Figure \ref{fig:thermal}, with the sublimation temperatures of H$_2$O, CO$_2$ and CO in the top panels. The remaining volatile species have been investigated further but, because of theoretical and/or observational arguments, none are thought to be the prime constituent of 1I/`Oumuamua. 

In  \citep{Seligman2020}, the hypothesis that 1I/`Oumuamua was composed of solid H$_2$ was explored. H$_2$ has a sublimation temperature of $\sim$6 K and this scenario suggested that 1I/`Oumuamua originated in a failed prestellar core of a giant molecular  \citep{fuglistaler2018}.  The continuous ablation of H$_2$ from the surface via cosmic ray exposure and solar photons would naturally produce the elongated shape. Other authors argued that the cosmic radiation would have destroyed a macroscopic H$_2$ body before it reached the Solar System \citep{Hoang2020,Phan2021}, although \citep{Levine2021_h2} showed that  km-scale progenitors could survive for timescales $\le100$ Myr. However, any theory must account for not only the existance of 1I/`Oumuamua, but also for the progenitor galactic population of similar-sized objects. This population is inferred to exist based on 1I/`Oumuamua's detection frequency, and is estimated to contain $\sim10^{26}$ objects (see discussion in subsection \ref{subsec:galactic_number}). The primary  caveat with the solid H$_2$ scenario is that the frigid temperature requirements for the prestellar cores would be extremely challenging to achieve ubiquitously throughout star forming regions to account for the progenitor  population of interstellar objects. 

It was subsequently hypothesized that  1I/`Oumuamua was composed of solid N$_2$ instead, implying that it was a remnant from an impactor event on an exo-Pluto analogue that generated significant fragments of solid nitrogen \citep{jackson20211i,desch20211i}. Given that such bodies are large enough to differentiate volatiles, this theory may also explain the lack of dust coma seen in the object. However, this would require a very high efficiency of ejection of nitrogen-rich material from planetary systems if  stellar populations were producing such objects uniformly -- exacerbated by the fact that that high velocity impactors do not generically produce fragments with sufficient kinetic energy to escape from the host body \citep{Levine2021}. However, some of these issues would be alleviated if only M stars produced 1I/`Oumuamua like interstellar objects and if the typical impacts were at indirect impact angles \citep{Desch2022}. If this theory proves correct, then it is possible that the discovery of 1I/`Oumuamua was an outlier event, and we can expect few or no similar discoveries in the future \citep{Levine2021}. 

In \citep{Seligman2021}, it was investigated whether CO outgassing could produce the nongravitational acceleration, despite the nondetection with  \textit{Spitzer}. While theoretically viable, this explanation requires a highly fine-tuned sporadic outgassing throughout the trajectory. 

  In \citep{Bergner2023} it was demonstrated that the crystallization of amorphous water ice would provide sufficient radiolytically-produced and entrapped H$_2$ to account for the observed nongravitational acceleration. Energetic processing of an H$_2$O-rich icy body irradiated by galactic cosmic rays would produce entrapped H$_2$, a process that has been demonstrated in laboratory experiments \citep{Bar-Nun1985,Sandford1993,Watanabe2000,Grieves2005,Zheng2006,Zheng2006b,Zheng2007}. This crystallization would occur in the absence of sublimation of the overall ice matrix, thereby  explaining the lack of dust coma observed. 
  
  This explanation requires the subsurface of 1I/`Oumuamua to reach sufficient temperatures to crystallize ice via annealing of the amorphous water matrix. Moreover, the required temperatures between 15-140 K \citep{Bar-Nun1988,Sandford1993, Zheng2006} must be reached  at sufficient depths within the interior to produce sufficient H$_2$ to power the acceleration, given a laboratory measured $\sim20-35\%$ yield of entrapped H$_2$ \citep{Sandford1993}. More realistic   H$_2$/H$_2$O 
 yields of tens of percent are predicted for the suburface meters of a porous cometary composition exposed to cosmic rays \citep{Maggiolo2020}, which can penetrate to depths of tens of meters  \citep{Gronoff2020}. 

Solutions to the thermal depths presented in \citep{Bergner2023} are shown in Figure \ref{fig:thermal}. Indeed, the subsurface meters reach sufficient temperatures to crystallize  amorphous ice without producing significant H$_2$O sublimation. This explanation would therefore partially explain the lack of observed dust  (Table \ref{table:production}). Without the bulk ice matrix sublimating, the release of entrapped H$_2$ within a few meters of the subsurface would not release dust trapped within the ice matrix. The lack of surface dust may be due to interactions with ambient gas in the interstellar medium which preferentially remove small dust grains from the surface of long-period comets and interstellar comets \cite{stern1990}.

\subsection{1I/`Oumuamua in the Context of Solar System Asteroids}

\subsubsection{Active Asteroids}\label{subsec:asteroids_active}

Small bodies throughout the Solar System have classically been categorized based on their volatile-driven activity. Comets were defined as icy objects that produced dusty comae and presumably spent little or no time in the inner Solar System. Meanwhile, asteroids lacked volatiles due to their closer proximity to the Sun which subjects them to prolonged and intense irradiation. However, recent advances have shown that this simple classification is not entirely accurate. Specifically, a subset of asteroids appear active and either have detectable dust activity or nongravitational accelerations potentially caused by outgassing \citep{Jewitt2012,Hsieh2017,Jewitt22_asteroid}. Within this context, the active asteroids are directly relevant to 1I/`Oumuamua which displayed a hybrid of cometary and asteroidal properties. 

The Main Belt Comets (MBCs) reside in the asteroid belt and  display low levels of cometary activity \citep{Hsieh2006}. The first MBC to be discovered was Comet 133P/(7968) Elst-Pizarro  \citep{Elst1996,Boehnhardt1996,Toth2000,Hsieh2004}.  Several other MBCs have since been identified, and  occurrence rates of $<1/500$ and $\sim 1/300$ have been inferred from surveys \citep{Sonnet2011,Bertini2011,Snodgrass2017,Ferellec2022}.

The cause of dust activity in these active asteroids is not completely clear. As discussed previously, outgassing induces nongravitational accelerations \citep{Whipple1950,Whipple1951} which have been detected on some active asteroids with related dust activity \citep{Hui2017}. However, other effects caused from impacts \citep{Snodgrass2010} and rotation  \citep{Jewitt2014}  can  produce dust activity in the absence of outgassing. 

Intriguing cases of active asteroids that do not exhibit obvious outgassing are  (3200) Phaethon and (101955) Bennu, the target of the  OSIRIS-REx mission. (3200) Phaethon was first identified to be active because of  its  association with the Geminid meteoroid stream \citep{Gustafson1989,Williams1993}. It produces  a small dust tail around perihelia with optically observed micron-sized dust production rates of $\sim 3$ kg s$^{-1}$ \citep{Jewitt2010,Jewitt2013,Li2013,Hui2017b}. However, this level of activity was incompatible with the flux from  the meteoroid stream. Therefore other effects such as   thermally induced stresses \citep{Jewitt2010}, sublimation of minerologically bound sodium \citep{Masiero2021}, rotation \citep{Ansdell2014,Nakano2020} and geometry  \citep{Hanus2016b,Taylor2019} have been used to explain the activity.  (101955) Bennu was  identified as active only when particle ejection events were measured by the OSIRIS-REx spacecraft  \citep{Lauretta2019,Hergenrother2019}. The measured mass  flux was $\dot{M}_{\rm Dust}\sim10^{-4}$ g s$^{-1}$ \citep{Hergenrother2020}. However,  the source of the  activity of (101955) Bennu remains unsolved \citep{Bottke2020,Molaro2020,Chesley2020}.  

  \begin{figure}%[h]
\begin{center}
       \includegraphics[scale=0.36,angle=0]{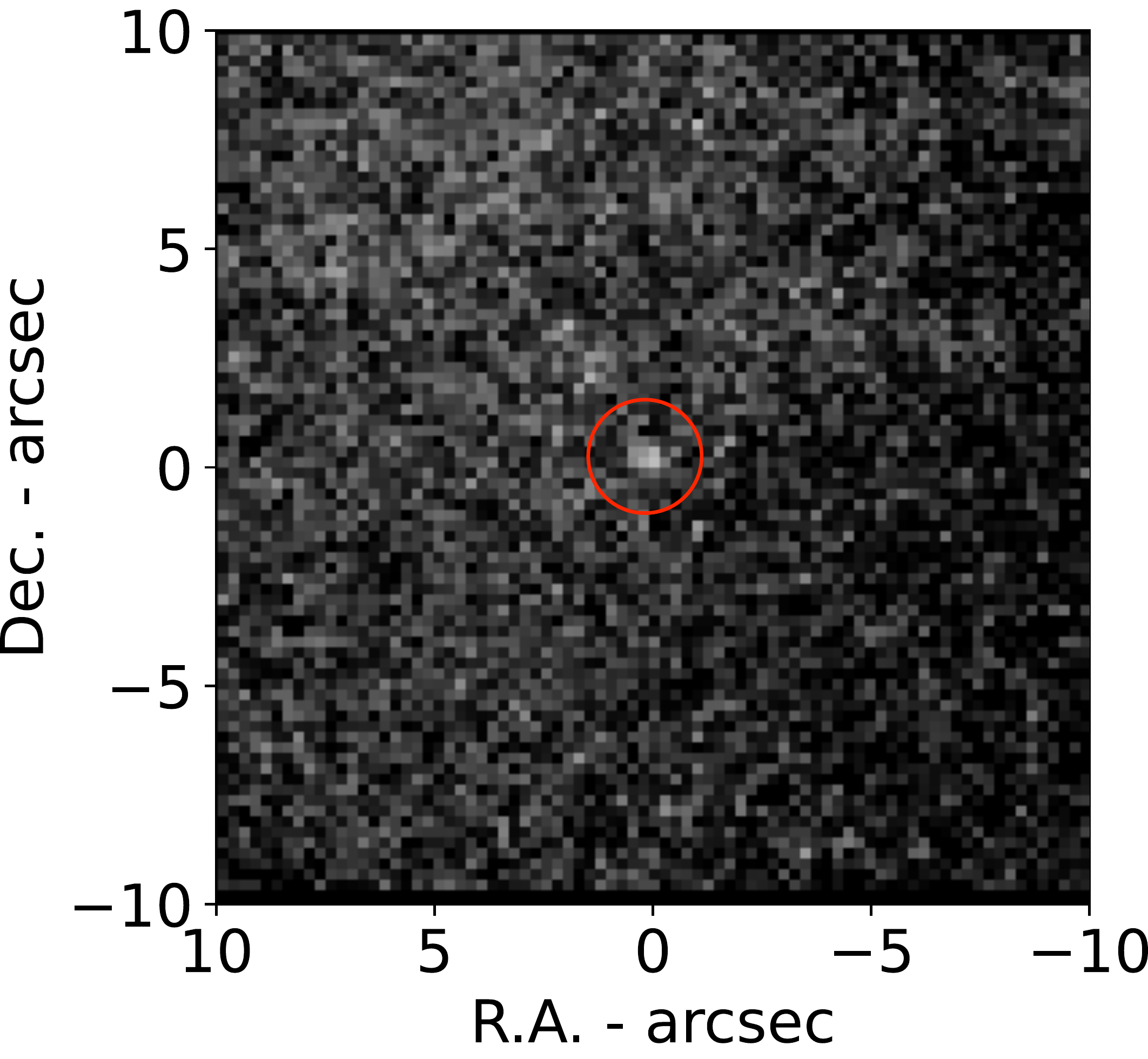}
    \caption{The dark comet  1998 KY$_{26}$ which exhibits no dust coma and a significant nongravitational acceleration. The image was taken with  the VLT and incorporates $\sim$ 1 hr of temporal coverage in 2020 December.  The $\sim30$ m object is  faint and indicated with a red circle.  Reproduced from \citep{Seligman2022}.} 
 \label{fig:ky26} 
\end{center}
\end{figure}

\subsubsection{Dark Comets (Inactive Asteroids Experiencing Nongravitational Accelerations)}\label{subsec:asteroids_darkcomets}
At the time of 1I/`Oumuamua's discovery, there had been no asteroidal objects discovered that exhibited nongravitational accelerations with no dust activity. As discussed previously, nongravitational accelerations have been measured on inactive asteroids due to the Yarkovsky effect and radiation pressure. However, the magnitudes of these accelerations are typically much lower than those discovered on 1I/`Oumuamua. Since the detection of 1I/`Oumuamua, a new population of inactive asteroids have been discovered that exhibit stronger nongravitational accelerations and similarly lack dust comae. We refer to these as dark comets.
 
Namely, \citep{Chesley2016}, \citep{Farnocchia2022} and \citep{Seligman2022} reported statistically significant detections of non-radial nongravitational accelerations in a sample of photometrically inactive  NEOs. The objects with nongravitational accelerations inferred from   astrometric data are  (523599) 2003 RM, 1998 KY$_{26}$, 2005 VL$_1$, 2016 NJ$_{33}$, 2010 VL$_{65}$, 2006 RH$_{120}$, and 2010 RF$_{12}$. In Figure \ref{fig:ky26}, we show a stack of  VLT images of one of these objects, 1998 KY$_{26}$, taken during 2020 December. This image contains a stack of exposures  resulting in a composite image with a total of 3600~s exposure.  The object quite clearly lacks a dust tail and, because of this and the nongravitational acceleration, is reminiscent of   1I/`Oumuamua. The image corresponds to an upper limit of dust production of $\dot{M}_{\rm Dust}<0.2$ g s$^{-1}$.

 \begin{figure*}%[h]
\begin{center}
       \includegraphics[scale=0.45,angle=0]{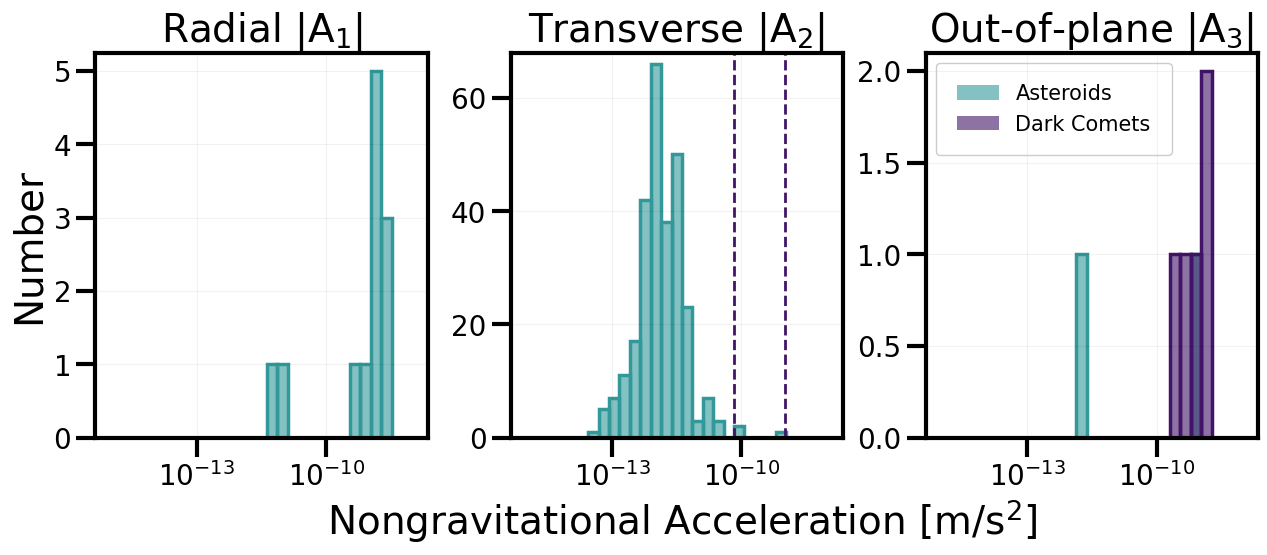}
    \caption{Nongravitational accelerations measured in asteroids (teal) and dark comets (purple), similar to Figure \ref{fig:nongravs}. Measured  radial nongravitational acceleration $A_1$ are due to solar radiation pressure (left) and  transverse acceleration $A_2$ are due to the Yarkovsky effect (middle).  The out-of-plane nongravitational acceleration $A_3$ measured in the objects (right) are presumably due to polar outgassing events.  Dark comet candidates with anomalous $A_2$ values, 2003 RM and 2006 RH$_{120}$, are shown in purple dashed lines in the middle panel. } 
 \label{fig:accelerations} 
\end{center}
\end{figure*}

 All of the dark comet nongravitational accelerations are inconsistent with being caused by radiation pressure or Yarkovsky effect, demonstrated in Figure \ref{fig:accelerations}. The best fitting nongravitational acceleration parameters and their significances are shown in Table \ref{table:objects}. The significant $A_2$ and $A_3$ are inconsistent in magnitude with radiation pressure and the Yarkovsky effect. Therefore \citep{Seligman2022,Farnocchia2022} argued that these accelerations were caused by outgassing with no (or low levels of) accompanying dust comae.

These dark comets may be Solar System analogues of 1I/'Oumuamua, in that they have measured nongravitational accelerations and no  (or low levels of)  dust comae. The lack of dust has been attributed to low levels of outgassing and presumably a lack of surface dust. Although 1I/`Oumuamua is now long gone, investigating these dark comets may reveal similar physical processes that can produce outgassing and nongravitational accelerations without significant dust production. The JAXA Hayabusa2 extended mission will rendezvous with 1998 KY$_{26}$ \citep{Hirabayashi2021}. The  instrument suite onboard Hayabusa2 can measure  volatile and dust production \citep{Kameda2015,Watanabe2017,Arai2017,Okada2017,Takita2017,Mizuno2017,Senshu2017,Yamada2017,Kameda2017,Iwata2017,Suzuki2018,Tatsumi2019}. Therefore, the nature of the nongravitational acceleration of these possible Solar System analogues of 1I/`Oumuamua should be definitively identified in the near-term future.

 \setlength{\tabcolsep}{1.5pt}
\begin{table*}[t]
\begin{center}   
\caption{Dark comets with  nongravitational accelerations and no observed dust production.   Best-fit nongravitational acceleration parameters for each object are shown with  1-$\sigma$ uncertainties. $P_{\rm Rot}$ is the rotational period and $H$ is the reported absolute magnitude. Reproduced from \citep{Seligman2022}.}
\label{table:objects}
\footnotesize
\begin{tabular}{lcccccccccc}
 Object  & $a$ & $e$ & $i$ & $q$ & $H$ & $r_{\rm n}$ & $P_{\rm Rot}$ & $A_1$ & $A_2$ & $A_3$\\
 & [au] &  & [$^\circ$] & [au] & [mag] &[m]& [h] & [$10^{-10}$ au d$^{-2}$] & [$10^{-10}$ au d$^{-2}$] & [$10^{-10}$ au d$^{-2}$]\\
  &  &  & &  & & &  & Signif.[$\sigma$] & Signif.[$\sigma$] & Signif.[$\sigma$]  \\
 \hline
 2003 RM & 2.92 & 0.60 & 10.86 & 1.17 & 19.80 & 230 &                & -1.045$\pm$1.217 &\phantom{-}0.0215 $\pm$0.0004   &\phantom{-}0.0156$\pm$0.0543 \\
  &  &  &  &  &  &  &  &$<1$ & 56 & $<1$  \\
1998 KY$_{26}$ & 1.23 & 0.20 & 1.48 & 0.98 & 25.60& 15& 0.178       &\phantom{-}1.73 $\pm$0.91   & -0.00126$\pm$0.00061     &\phantom{-}0.320  $\pm$0.115 \\
  &  &  &  &  &  &  &  & 2 & 2 & 3\\
2005 VL$_1$ & 0.89 & 0.23 & 0.25 & 0.69 & 26.45 & 11 &              & -6.66 $\pm$8.02 & -0.00711$\pm$0.00592    & -0.240 $\pm$0.041    
\\
  &  &  &  &  &  &  &  &$<1$ & 1 &6    
\\
2006 RH$_{120}$& 1.00 & 0.04  & 0.31  & 0.96 & 29.50 & 2-7 & 0.046 &\phantom{-}1.38 $\pm$0.08    & -0.507  $\pm$0.0637   & -0.130 $\pm$0.032   \\
  &  &  &  &  &  &  &  &18   &8  & 4    \\
2010 VL$_{65}$ & 1.07 & 0.14 & 4.41 &0.91 & 29.22&3&                &\phantom{-}6.57 $\pm$13.0  & -0.00146$\pm$0.00534  & -0.913 $\pm$0.130     \\ 
  &  &  &  &  &  &  &  &$<1$ &$<1$ & 7      \\ 
2010 RF$_{12}$&1.06 & 0.19  & 0.88  &0.86 &28.42 &4 &               &\phantom{-}0.488$\pm$0.597 & -0.00136$\pm$0.00286  & -0.168 $\pm$0.021    \\
  &  &  &  &  &  &  &  &$<1$ & $<1$ & 8    \\ 
  2016 NJ$_{33}$ & 1.31 & 0.21 & 6.64 & 1.04 & 25.49&16&$0.41$-$1.99$ &\phantom{-}9.28 $\pm$2.96 &\phantom{-}0.00566$\pm$0.00193     &\phantom{-}0.848  $\pm$0.163     \\
  &  &  &  &  &  &  &  &3 & 3 &5    \\
\hline
\end{tabular}
\end{center}
\end{table*}

\section{2I/Borisov}

\begin{figure}
\begin{center}
       \includegraphics[scale=0.5,angle=0]{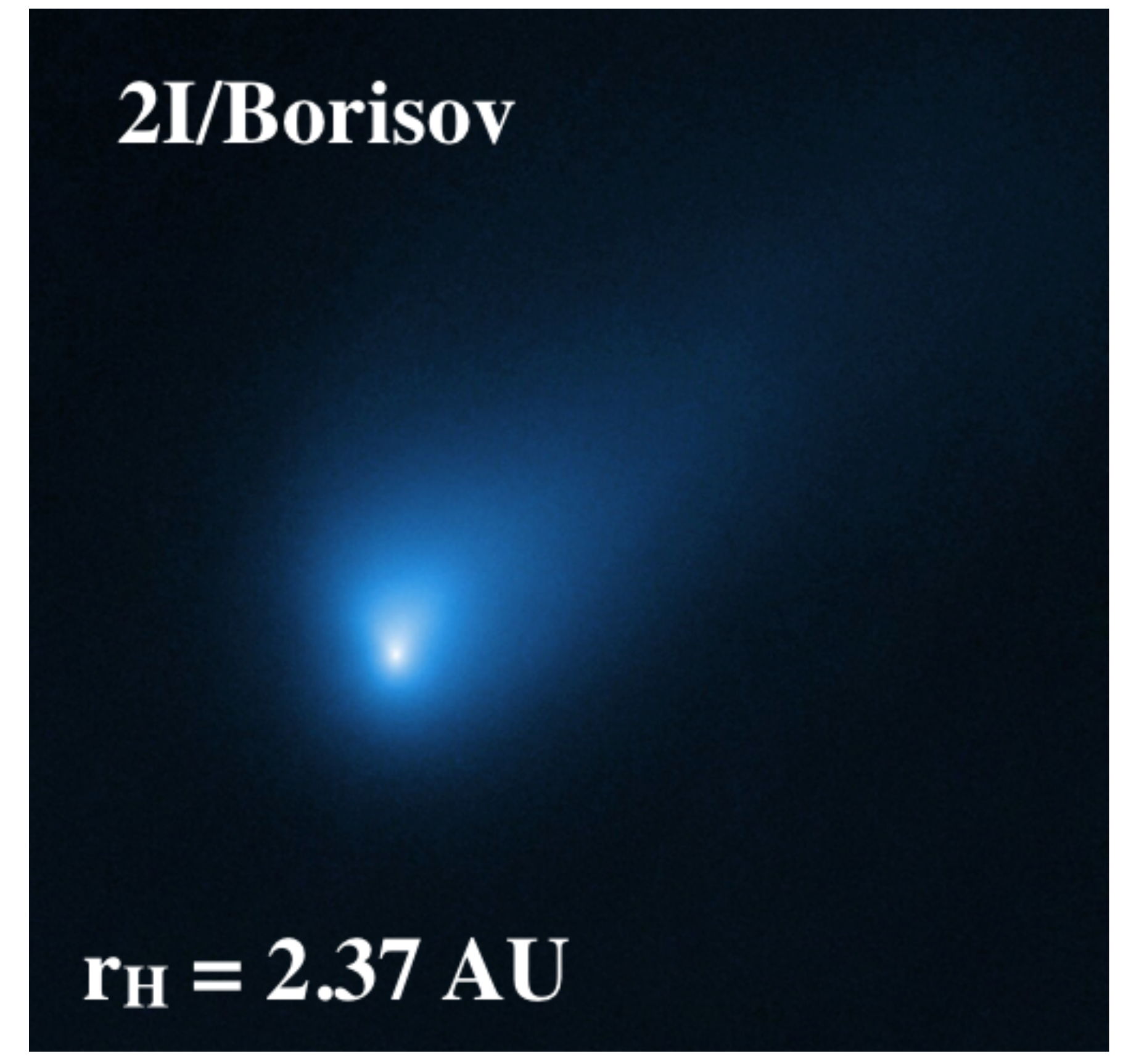}
    \caption{ The second interstellar comet discovered, 2I/Borisov. The image was taken with  HST  on 2019 October 12  \citep{Jewitt20}. 2I/Borisov  displays a clear cometary tail.   $r_H$ corresponds to the heliocentric distance of the object when the image was obtained. Reproduced from \citep{Jewitt20}.}
 \label{Fig:2I} 
\end{center}
\end{figure}

\subsection{Discovery}

The  comet C/2019 Q4 (Borisov) was discovered on   2019 August 30  when it was  close to the sun with an angular separation of just 38$^{\circ}$. This object was discovered by Gennadiy Borisov, an amateur astronomer using a 0.65 m telescope. It is worth noting that the object was not discovered by the all sky surveys searching for transient events, which intentionally do not observe near the Sun.  

Like 1I/`Oumuamua, the trajectory of the comet is definitively of interstellar origin, with an escape velocity, $v_{\infty} = 32.304 \pm 0.001$ km/s, semimajor axis of $a=-0.850$ au, eccentricity of $e=3.363$, inclination $i=44.0^\circ$ and perihelion of $q=2.009$ au. Because  of the definitive interstellar trajectory, it was renamed to 2I/Borisov. Unlike 1I/`Oumuamua, the object was discovered $\sim3$ months prior to perihelia which left ample time for characterization with subsequent telescopic observations.  In Figure \ref{Fig:2I}, we show an image of 2I/Borisov taken with the \textit{Hubble Space Telescope} \citep{Jewitt20}.

\subsection{Nucleus Properties}

\subsubsection{Size}

The  images of 2I/Borisov obtained  with HST \citep{Jewitt2019b} provided empirical evidence for a nuclear radius of 0.2--0.5 km. This was supported by upper limits from ground-based infrared imaging \citep{Bolin2019}. The upper limit was derived from modelling the brightness profile of the coma, and the lower limit was derived from the measured nongravitational acceleration assuming a bulk density representative of comets to prevent catastrophic disruption due to outgassing forces \citep{Jewitt20}. This nuclear size is more in line with Solar System comets, and provides a stark contrast to the very small size inferred for 1I/`Oumuamua.

\subsubsection{Colors}
2I/Borisov  displayed a distinct dust and gas coma and was unmistakably a comet \citep{Jewitt2019b,Bolin2019,Guzik:2020,Hui2020,Mazzotta21}. The measured colors of 2I/Borisov, unlike for 1I/`Oumuamua, were representative of the color of the ejected dust which dominate the images. The dust ejected from 2I/Borisov was reddened with respect to that of the Sun, similar to 1I/`Oumuamua.   Early spectral observations with the 10.4 m Gran Telescopio Canarias (GTC) revealed a visible spectrum  with a spectral slope that is roughly in the middle of the range of visible spectral slopes observed for cometary nuclei in the Solar System \citep{deleon2019}. The dust in the cometary tail was measured to have a slightly reddish color of $g'-r' = 0.63 \pm 0.03$ \citep{Guzik:2020,Jewitt2019b, Bolin2019}, later measured as $g'-r' = 0.68 \pm 0.04$ and $r'-i' = 0.23 \pm 0.03$ \citep{Hui2020}.

\subsubsection{Nongravitational Acceleration}

\begin{figure}%[h]
\begin{center}
       \includegraphics[scale=0.42,angle=0]{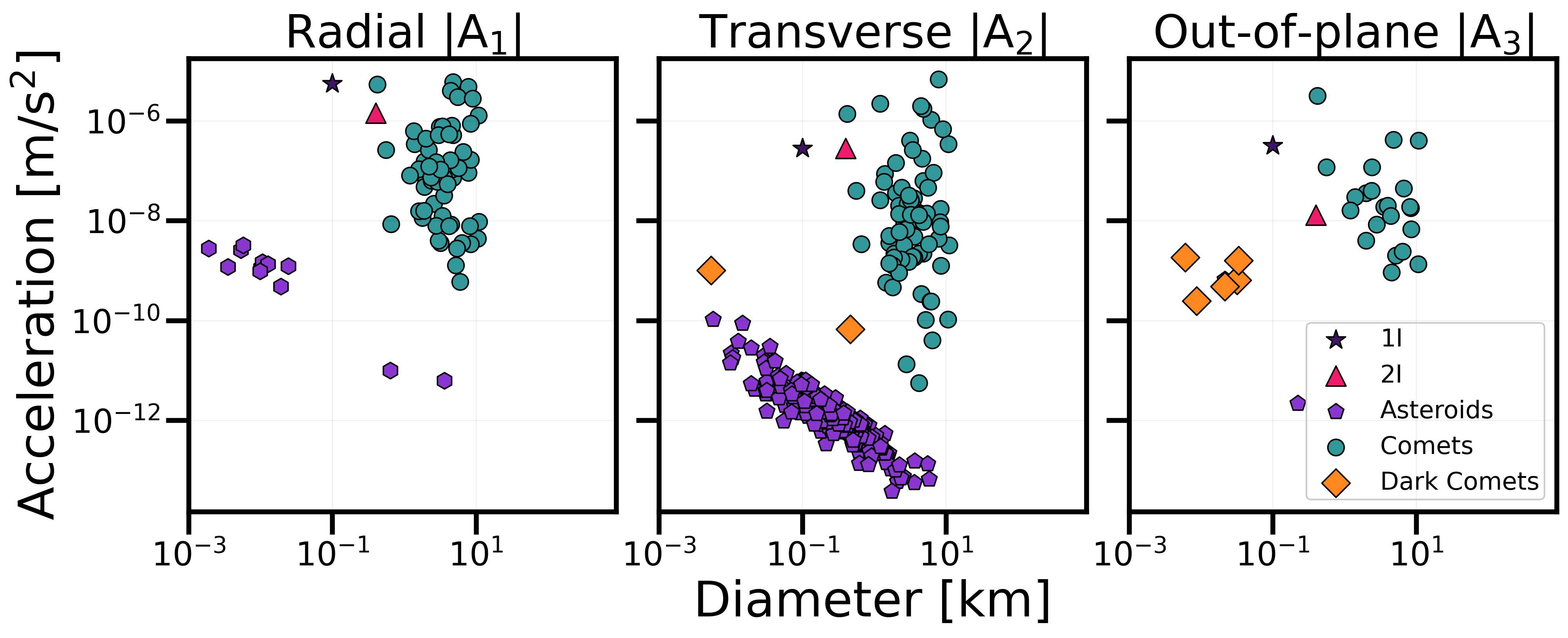}
    \caption{Measured nongravitational accelerations of a variety of small bodies. The three panels  show measured components of nongravitational accelerations in the radial A$_1$ direction (left), transverse A$_2$ direction (middle) and out-of-plane A$_3$ direction (right). In each panel, measured nongravitational accelerations of comets (teal circles) asteroids (purple hexagons), dark comets (orange diamonds), 1I/`Oumuamua (dark purple star) and 2I/Borisov (magenta triangle) are shown. Only comets with measured nuclear diameters are shown --- many more comets exist for which nongravitational accelerations have been measured (Figure \ref{fig:nongravs}). The size of the asteroids have been approximated using Equation \ref{eq:diameter_mag}.}\label{fig:all_nongravs} 
\end{center}
\end{figure}

With 10 months of astrometric positional data, it was evident that 2I/Borisov exhibited nongravitational acceleration. Similar to 1I/`Oumuamua,  the dominant component of the nongravitational acceleration was in the radial direction, with $A_1 = 7.1\pm0.8 \times 10^8$ au d$^{-2}$, $A_2 = -1.4\pm0.3 \times 10^8$ au d$^{-2}$ and $A_3 = 0.1\pm1.5 \times 10^8$ au d$^{-2}$. These   were consistent with being recoil effects from measured production rates \citep{Hui2020,delafuente2020,Manzini2020}.  In Figure \ref{fig:all_nongravs}, we show the nongravitational accelerations measured in comets, asteroids, dark comets,  1I/`Oumuamua and 2I/Borisov. We limit to only objects that have sizes measured.

\subsection{Coma Properties}

\subsubsection{Dust Properties}

VLT and ALMA data of 2I/Borisov was reported by \citep{yang2021} showing that that the dust in the coma contained a contribution from pebbles with sizes $\gtrsim1$ mm. Dust production rates of $Q({\rm dust})=35$ kg  s$^{-1}$ (2019 November) and $Q({\rm dust})=30$ kg  s$^{-1}$ (2019 December) were reported by \citep{Cremonese2020} using observations with the Telescopio Nazionale Galileo. The contributing dust grains were consistent with sizes mostly larger than $100\,\mu \text{m}$ \citep{Kim2020}.

The presence of submicron sized conglomerates of magnesium-ferrous particles were inferred from photometric observations for $\sim2$ months before periastron   \citep{Busarev2021}. Polarimetric measurements of the object obtained by the ESO Very Large Telescope intriguingly showed that the dust in the outflow exhibited a high level of polarization compared to levels seen in dust produced in typical Solar System comets \citep{Bagnulo2021}.  This polarization was further shown to be consistent with a high $\sim80\%$ fraction of rough fractal aggregates and $\sim20\%$ of agglomerated debris, interpreted as 2I/Borisov having a large percentage of small and porous \textit{pristine} cosmic dust particles \citep{Halder2023}.

\subsubsection{Volatile Composition}

  \begin{table}[h]
   
\tabcolsep7.6pt
\caption{The production rates of molecules CO, H$_2$O and OH measured in the comae of 2I/Borisov. r$_H$ is the Heliocentric distance at observation. Adapted from \cite{Seligman2022PSJ}.}
\label{Table:waterco}
\begin{center}
%\begin{tabular}{@{}l|c|c|c|c|c@{}}
\begin{tabular}{@{}lccccc@{}}

Date & r$_H$  & ${\rm Q({H_2O})}$ $10^{26}$ & ${\rm Q({CO})}$$10^{26}$&  ${\rm Q({OH})}$$10^{26}$ & Reference \\
&[au]& [molec s$^{-1}$]&[molec s$^{-1}$]&[molec s$^{-1}$]&\\
\hline
9/27/19&2.56&$<8.2$&&&\cite{Xing2020} \\
10/2/19&2.50&&&$<0.2$&\cite{Opitom:2019-borisov}\\
10/11/19&2.38&$6.3\pm1.5$&&&\cite{McKay2020} \\
10/13/19&2.36&&&$<0.2$&\cite{Opitom:2019-borisov}\\
11/1/19&2.17&$7.0\pm1.5$&&&\cite{Xing2020}\\
12/1/19&2.01&$10.7\pm1.2$&&&\cite{Xing2020}\\
12/3/19&2.01& &$3.3\pm0.8$&&\cite{yang2021}\\
12/11/19&2.01&&$7.5\pm2.3$&&\cite{Bodewits2020}\\
12/15-16/19&2.02&&$4.4\pm0.7$&&\cite{Cordiner2020}\\
12/19-22/19&2.03 &$4.9\pm0.9$&$6.4\pm1.4$&&\cite{Bodewits2020}\\
12/21/19&2.03 &$4.9\pm0.9$&&&\cite{Xing2020}\\
12/30/19&2.07&&$10.7\pm6.4$&&\cite{Bodewits2020}\\
1/13/20&2.16&$<5.6$&$8.7\pm3.1$&&\cite{Bodewits2020} \\
1/14/20&2.17&$<6.2$&&&\cite{Xing2020}\\
2/17/20&$2.54$&$<2.3$&&&\cite{Xing2020}\\\hline
\end{tabular}
\end{center}
\end{table}

      \begin{table}[h]
\tabcolsep7.6pt
\caption{The production rates of molecules CN, C$_2$ and C$_3$ measured in the coma of 2I/Borisov. r$_H$ is the Heliocentric distance at observation. Adapted from \cite{Seligman2022PSJ}. }
\label{Table:otherprods}
\begin{center}
\begin{tabular}{@{}lccccc@{}}
Date & r$_H$  & ${\rm Q({CN})}$ & ${\rm Q({C_2})}$ & ${\rm Q({C_3})}$ &  Reference \\
&[au]& $10^{24}$[molec s$^{-1}$]&$10^{24}$[molec s$^{-1}$]&$10^{24}$[molec s$^{-1}$]&\\
\hline

9/20/19&2.67&$3.7\pm0.4$&$<4$&&\cite{Fitzsimmons:2019}\\
9/20/19&2.67&$<5$&$<8$&&\cite{Kareta:2019}\\

10/1/19&2.50&$1.1\pm2.0$&$<2.5$&&\cite{Kareta:2019}\\
10/1/19&2.51&$1.8\pm0.1$&$<0.9$&$<0.3$& \cite{Opitom:2019-borisov}\\
10/2/19&2.50&$1.9\pm0.1$&$<0.6$&$<0.2$&\cite{Opitom:2019-borisov}\\
10/9/19&2.41&$1.59\pm0.09$&$<0.44$&& \cite{Kareta:2019}\\
10/10/19&2.39&$1.69\pm0.04$&$<0.162$&&\cite{Kareta:2019}\\

10/13/19&2.36&$2.1\pm0.1$&$<0.6$&$<0.3$&\cite{Opitom:2019-borisov}\\
10/18/19&2.31&$1.9\pm0.6$&&&\cite{Opitom:2019-borisov}\\
10/20/19&2.29&$1.6\pm0.5$&&&\cite{Opitom:2019-borisov}\\
10/26/19&2.23&$1.9\pm0.3$&&&\cite{Kareta:2019}\\
10/31/19&2.18&$2.0\pm0.2$&&&\cite{Lin2020}\\
11/4/19&2.15&$2.4\pm0.2$&$0.55\pm0.04$&$0.03\pm0.01$& \cite{Lin2020}\\
11/10/19&2.12&$1.9\pm0.5$&&& \cite{Bannister2020}\\
11/14/19&2.09&$1.8\pm0.2$&$1.1$&& \cite{Bannister2020}\\
11/17/19&2.08&$1.9\pm0.5$&&&\cite{Bannister2020}\\
11/25/19&2.04&$1.6\pm0.5$&&&\cite{Bannister2020}\\
11/26/19&2.04&$1.8\pm0.2$&&& \cite{Bannister2020}\\
11/26/19&2.04&$1.5\pm0.5$&$1.1$&& \cite{Bannister2020}\\
11/30/19&2.01&$3.36\pm0.25$&$1.82\pm0.6$&$0.197\pm0.052$&\cite{Aravind2021}\\
12/22/19&2.03&$6.68\pm0.27$&$2.3\pm0.82$&$0.714\pm0.074$& \cite{Aravind2021}\\\hline
\end{tabular}
\end{center}
\end{table}

The orbit and discovery of 2I/Borisov enabled several months of detailed compositional measurements  to be obtained.   While the object was definitively a comet unlike 1I/`Oumuamua, its composition was different than those typically seen in Solar System comets. These observations are summarized in Table \ref{Table:waterco} and Table \ref{Table:otherprods}. 

Initial measurements of the object showed that ${\rm H_2O}$ was active as early as 6 au and inbound \citep{Fitzsimmons:2019, Jewitt2019b, Ye:2019}. Subsequent observations revealed carbon and nitrogen bearing species in the coma. Near-UV emission of CN from  2019 September 20 was presented by  \citep{Fitzsimmons:2019} presented, together with an upper limit on the abundance of $\rm C_2$, based on observations with the 4.2-meter William Herschel Telescope and the ISIS spectrograph on La Palma.  Measurements of CN and $\rm C_2$ production were also reported by \citep{Kareta:2019} using the 2.3-meter Bok telescope at Kitt Peak National Observatory in Arizona, the 6.5-meter MMT telescope located at Mount Hopkins, and the Large Binocular Telescope at the DDT. Similar spectroscopic detections and upper limits were reported by \citep{Opitom:2019-borisov} using data from the 4.2-meter William Herschel and 2.5-meter Isaac Newton telescopes on a range of dates from 2019 September 30 to 2019 October 13. Detections of $\rm C_2$ and  of $\rm CN$ were reported by \citep{Lin:2019} using data from the MDM observatory Hiltner 2.4-meter telescope and the Ohio State Multi-Object Spectrograph from 2019 October 31 and November 4. Additional observations were presented by \citep{Bannister2020} using the Multi-Unit Spectroscopic Explorer (MUSE) at the 8.2-meter UT4 of the ESO/Very Large Telescope (VLT) and the the 0.6-meter TRAPPIST North and South telescopes.

 Production rates and upper limits of H$_2$O were measured throughout the trajectory. A production rates of $Q({\rm H_2O})=6.3\pm1.5\times 10^{26}$ molecules  s$^{-1}$ was derived by \citep{McKay2020} based on observations with the ARCES instrument at Apache Point Observatory on 2019 October 11, while \citep{Aravind2021} reported observations taken with the 2-meter Himalayan Chandra Telescope located at the Indian Astronomical Observatory, Hanle (HCT) and the Mount Abu Infrared Observatory (MIRO) from 2019 November 30 and December 22 that yielded production rate ratios of $Q(C2)/Q(CN) = 0.54 \pm 0.18$ and  $Q(C2)/Q(CN) = 0.34 \pm 0.12$ before and after perihelion respectively. Critically, \citep{Xing2020} presented observations of production rates both before and after perihelion with 6 epochs of observations with the Neil Gehrels Swift Observatory's Ultraviolet/Optical Telescope. These revealed that the H$_2$O production increased and decreased before and after perihelion, and they estimated an active fraction of $\ge$ 55 $\%$ of the surface to account for the inferred production.

While these previously discussed observations were in line with compositions typically seen in Solar System comets, post perihelia observations demonstrated that the object was enriched in the hypervolatile CO \citep{Cordiner2020,Bodewits2020}. Atacama Large Millimeter/submillimeter Array (ALMA) images were obtained on 2019 December 15 and 16,  and revealed the presence of HCN and CO at high abundances  relative to H$_2$O \citep{Cordiner2020}. Similarly, \citep{Bodewits2020} reported observations with the Cosmic Origins Spectrograph (COS) on the HST between 2019 December 11  and 2020 January 13,  that revealed high CO production rates with respect to H$_2$O. Additional measurements of CO production by \citep{yang2021} found that the CO/H$_2$O mixing ratio changed drastically before and after perihelion. 
 
A spectroscopic detection of atomic nickel vapor in the coma of 2I/Borisov at 2.322 au was reported by \citep{Guzik2021} using observations with X-shooter spectrograph at the ESO VLT on 2020 January 28, 30 and 31, when the object was at an equilbrium temperature $\sim180$ K. Atomic nickel vapor has been detected in  sun-grazing comets such as the case of C/1965 S1 (Ikeya–Seki) \citep{Preston1967,Slaughter1969}.  However, this is typically observed at much warmer temperatures $>700$ K and attributed to the sublimation of metal enriched dust grains. These authors concluded that the nickel vapor was a photodissaociation product of a short-lived nickel-containing molecule.

\subsection{Breakup}

2I/Borisov exhibited a stark brightening event \citep{Drahus2020ATel} and subsequent breakup in the spring of 2020 \citep{Jewitt2020:BorisovBreakup, Jewitt2020ATel, Bolin2020ATel,Zhang2020ATel}.  HST observations of the breakup were presented in \citep{Jewitt2020:BorisovBreakup}. The splitting event can be seen in March, where the nucleus develops  a two-lobed  shape. 

The evolution of the coma morphology of 2I/Borisov in the  HST images was interpreted   as evidence of seasonal effects \citep{Kim2020}. Specifically, the changes in activity levels and breakup \citep{Jewitt2020:BorisovBreakup} could be explained by regions on the northern hemisphere of the nucleus being exposed to the sun for the first time (particularly, see Figure 7 in \citep{Kim2020}). Seasonal effects and nucleus disruption are typically seen in Solar System comets and have been attributed to cometary fading \citep{Brasser2015}.

\subsection{Borisov in the Context of Solar System Comets}

\begin{figure}%[h]
\begin{center}
       \includegraphics[scale=0.35,angle=0]{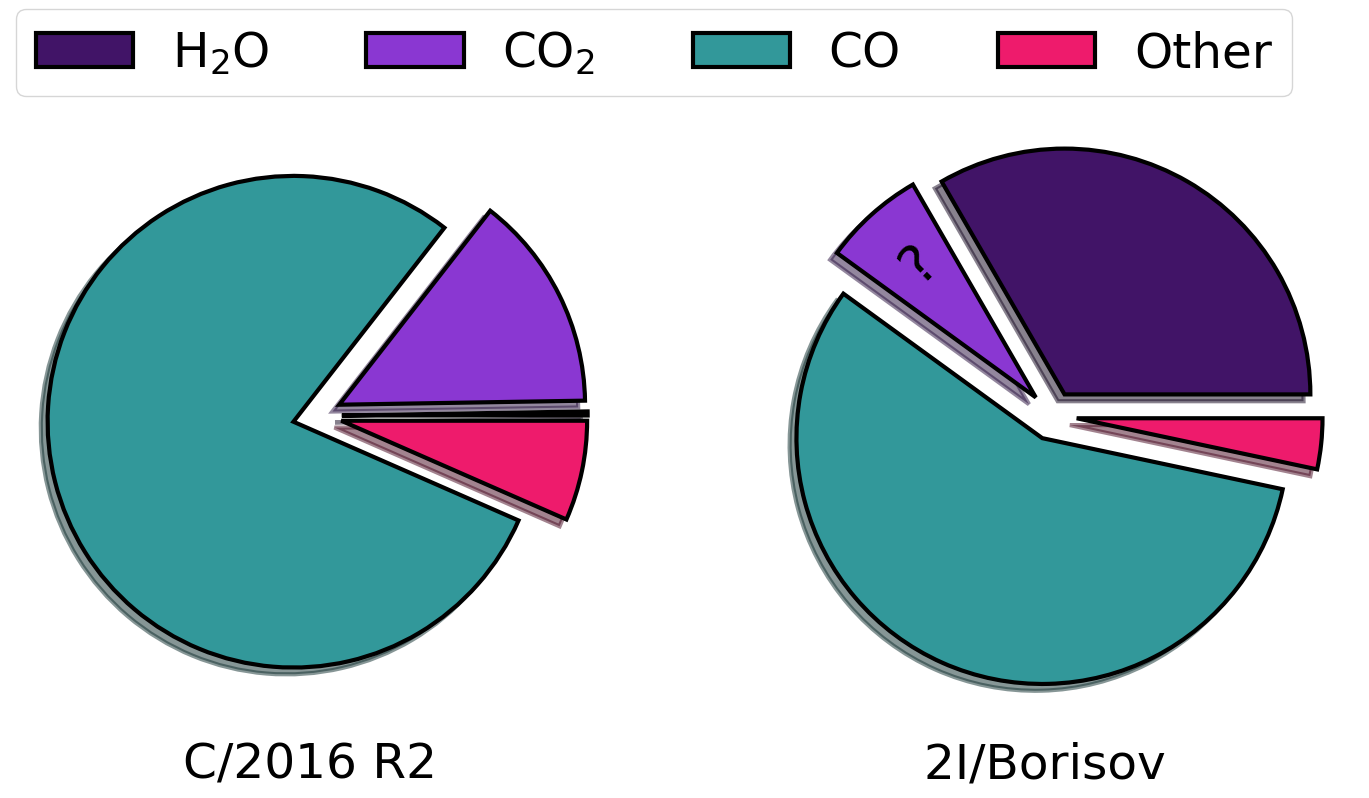}
    \caption{The compositions of 2I/Borisov and C/2016 R2, analogous to Figure \ref{fig:Piecharts}. Note that there were no observations sensitive to CO$_2$  obtained for 2I/Borisov. Therefore, it is possible that the object had CO$_2$ in its outflow that was never measured as indicated with a question mark.   }\label{fig:Piecharts_2I} 
\end{center}
\end{figure}

The composition of 2I/Borisov was anomalous compared to typical Solar System comets due to the high abundance of the hypervolatile CO. Contemporaneous observations sensitive to H$_2$O and CO revealed a production rate ratio of $Q({\rm CO}) / Q({\rm H_2O}) \ge$ 1. It is important to note that this production rate ratio is calculated only when near-contemporaneous measurements of both species were obtained. As discussed previously in this review (subsection \ref{subsec:comets_outgassing} and \ref{subsec:comets_compositions}), comets are typically composed primarily of H$_2$O, with average ratios of $Q({\rm CO}) / Q({\rm H_2O}) \sim$ 4$\%$ although with a wide range \citep{Bockelee22}.

This was interpreted as 2I/Borisov having  formed in a carbon rich environment. It was argued that  2I/Borisov likely formed in an M star system. In such a system, the relatively cooler temperatures would lead to a CO snowline closer to the star. This implies that the majority of the circumstellar solids lie outside the CO snowline \citep{Bodewits2020}. Alternatively, \citep{Cordiner2020} argued that 2I/Borisov formed close to or at the CO snowline and was possibly the remnant of an impacted exo-Kuiper belt object.  This scenario is similar to that suggested for 1I/`Oumuamua in which it is a fragment of an exo-Pluto analogue \citep{desch20211i,jackson20211i,Desch2022}. All of these interpretations of the hyerpvolatile content indicate that the object formed exterior to the CO snowline in its original protoplanetary disk \citep{Price2021}. 

Interestingly, the abundance of remnant hypervolatiles can inform the timing of the ejection of 2I/Borisov from its host system. Even if it formed beyond the CO snowline and remained there, prolonged heating from stellar irradiation would have removed hypervolatiles  at large stellocentric distances if exposed for 100s of Myr. Therefore, \citep{Lisse2022} argued that the object must have been  ejected within $<20$ Myr of the formation in its host system. 

An  intriguing comparison object is C/2016 R2 \citep{Wierzchos2018, Cochran2018,McKay2019}, a highly eccentric  LPC that exhibits a CO to H$_2$O production rate ratio $\gtrsim100$ \citep{McKay2019}. However, C/2016 R2 displayed a  CO/H$_2$O ratio much higher than that of 2I/Borisov, and an unexpectedly high abundance of N$_2$ \citep{Wierzchos2018,Mousis2021}. It has been suggested that C/2016 R2 is interstellar in origin because of its exotic composition \citep{McKay2019}. In Figure \ref{fig:Piecharts_2I}, we show the compositions of 2I/Borisov and C/2016 R2.  While not a direct analogue of 2I/Borisov, it is one of the few Solar System comets enriched in hypervolatiles, and is certainly the most extreme case. Relevant to this discussion, the Solar System comets C/2009 P1 Garradd \citep{Paganini2012,McKay2015,Feaga2014,Bodewits_2014}, C/2010 G2 Hill  \citep{Kawakita2014} and C/2006 W3 Christensen \citep{Ootsubo2012}  have high abundances of CO relative to H$_2$O.  C/1995 O1 (Hale-Bopp) had $Q({\rm CO}) / Q({\rm H_2O}) >$ 12 measured at $r_H$ = 6 au \citep{Biver02} and the  Centaur 29P/SW1 exhibited $Q({\rm CO}) / Q({\rm H_2O}) =$ 10$\pm$1, measured at $r_H \sim$ 6 au \citep{Biver02,Bockelee22}). However, at these further heliocentric distances, H$_2$O is typically not active. Therefore, the  compositions measured at large distances are not necessarily representative of the volatile inventories.

\section{Galactic Population of Interstellar Objects}\label{GalacticPop}

\subsection{Number of Interstellar Objects}\label{subsec:galactic_number}
Prior to the detection of 1I/`Oumuamua, several studies had estimated the number density of interstellar comets from non-detections. These are summarized in Table \ref{Table:galactic_pop}, dating back to 1976. Immediately before 1I/`Oumuamua's discovery, several studies provided detailed simulations of the expected galactic population of ejected comets and predicted the prospects for future detections with existing and forthcoming surveys. 

The discovery of 1I/`Oumuamua implies that a much larger galactic population exists than  was previously estimated. The  occurrence rate of detections was estimated  to be $0.2$ yr$^{-1}$, with an average of $\sim 20$ $M_{\oplus}$ of material ejected per star \citep{Trilling2017}.  The  number density was estimated to be $n \sim 10^{-2} \, {\rm au}^{-3}$ \citep{Laughlin2017}, implying a total number of $N\sim 2\times 10^{26}$ throughout the Milky Way and a galactic mass of $M\sim10^{11}\,M_\oplus$. A more detailed analysis that incorporated the survey sensitivities of observational facilities  resulted in a number density  $n \sim2\times 10^{-1}\,{\rm au}^{-3}$ \citep{Do2018}.  If we assume a mass of $M_{\rm 1I}=10^{12}\,\rm{g}$ in every 1I/`Oumuamua-like object, this  implies a total galactic mass density of 0.3 $M_{\oplus}\,{\rm pc}^{-3}$. This translates to   a population of $N\sim4\times10^{26}$ in the Milky Way, or approximately $1\,M_{\oplus}$ per stellar system. Based on the non-detection of similar objects since these initial estimates, this number density has been revised to $n\sim1\times 10^{-1}\,{\rm au}^{-3}$ \citep{Levine2021,Jewitt2022ARAA}. Frustratingly, the detection of 2I/Borisov does not  constrain  the number density of interstellar objects due to its brightness and discovery method.

\begin{table}%[h]
\tabcolsep7.5pt
\caption{Published estimates of the galactic space density of interstellar interlopers. 
}
\label{Table:galactic_pop}
\begin{center}
\begin{tabular}{@{}lccc@{}}
Reference   & Density &Density & Rate \\
&[$M_\odot$ pc$^{-3}$]& [ au$^{-3}$  ]& [ yr$^{-1}$  ]\\

\hline
\citep{sekanina1976probability} & $6\times10^{-4}$&$\lesssim0.07^{\rm a}$&\\
\citep{mcglynn1989}             &2$\times10^{-5}$ &10$^{-3}\,^{\rm a}$&$\sim$0.6\\
\citep{Sen1993}                 & & &$<$0.01\\
\citep{Jewitt2003}              &&$ \sim 10^{-3}$\,$^{\rm a}$&\\
\cite{francis2005}              & $3\times10^{-6}$ & $3-4.5\times10^{-4}$&\\
 \citep{Moro2009}               & $\sim2\times10^{-7}$ & 3$\times10^{-5}$\,$^{\rm a}$&\\
 \citep{Engelhardt2014}         & & $<1.4 \times 10^{-4}$ & \\
 \citep{Cook2016}               & & 10$^{-8}$ - 10$^{-5}\,^{\rm a}$ & \\
 \citep{Jewitt2017}             & & $\sim$ 0.1&\\
  \citep{Trilling2017}          & & $\sim$ 0.1&\\
   \citep{Laughlin2017}         & & $\sim$ 0.2&\\
  \citep{Do2018}                & & $\sim$ 0.2&\\
  \citep{MoroMartin2018i}& $\sim3\times10^{-7}$ ({\rm singles}) & & \\
 \citep{MoroMartin2018i}& $\sim1\times10^{-6}$ ({\rm binaries}) & & \\
  \citep{Moro2019exOC}   &  & $\sim3\times10^{-3}$ (Oort cloud)&\\
  \hline
\end{tabular}
\end{center}
\begin{tabnote}
$^{\rm a}$ Objects greater than 1 km.
\end{tabnote}
\end{table}

As mentioned above, the number density of interstellar objects is higher than expected, if we assume that 1I/‘Oumuamua is representative of a population that is uniformly distributed. Population studies that have tackled this discrepancy have estimated the number density of interstellar planetesimals based on the observed number of stars per unit volume of space and an estimated number of ejected planetesimals per star.

\subsubsection{Contribution from Protoplanetary Disks Around Single Stars and Close Binaries}

 Feasible sources of planetesimal ejection are dynamical instabilities and orbit readjustment that occur as a result of interactions between planetesimal and growing planets. These interactions lead to planetary migration, as  is thought to have occurred in the early Solar System \citep{Gomes2005Nat,Nesvorny2018}. Numerical simulations show that these episodes of dynamical instability and planetesimal ejection are common in many planetary configurations \citep{Rayond2018MNRAS}. In the Solar System, there is evidence that the planetesimal belts were heavily depleted and that the primordial Kuiper and asteroid belts were significantly more massive. The two pieces of evidence are the existence of Kuiper belt objects larger than 200 km (which formation by pairwise accretion must have required a number density of objects about two orders of magnitude higher than today), and the strong depletion in the asteroid belt region with respect to the minimum mass solar nebula. 

Ejection occurs when the giant planets in the Solar System have close encounters with planetesimals. Some of these close encounters would  impart sufficient energy to eject the planetesimal. Based on theoretical models of the timing of giant planet migration and/or instability in the Solar System, it seems likely that this  occurred within the first $ <10$ Myr  after the disk dispersed \citep{Grav2011,Buie2015,Nesvorny2018b,Clement2018,Clement2019,deSousa2019,Nesvorny2021,Morgan2021,Liu2022}. Moreover, the recent report of an excess of free-floating planets in the newly formed ($<10$ Myr) Upper Scorpius  stellar association \citep{Miret-Roig2022} is  consistent with ubiquitous early ejection of debris. 

For a giant planet to eject an interstellar comet, the planet must impart sufficient energy via a close encounter to scatter the comet onto an unbound trajectory. 
The Safronov number $\Theta = V_e^2/(2V_K^2)$  is useful to quantify the efficiency of this process, where $V_e$ is the escape velocity and $V_K$ is the orbital velocity.   Only objects with $\Theta>1$ are capable of ejecting objects. The Safronov number may be written as, 
\begin{equation}\label{eq:safronov}
  \Theta= \left(\frac{M_{\rm P}}{M_*}\right) \left(\frac{a_{\rm P}}{R_{\rm P}}\right) \,.
\end{equation}
In Equation \ref{eq:safronov}, $M_*$ and $M_{\rm P}$ are the mass of the star and planet, while  $a_{\rm P}$ and $R_{\rm P}$ are the semimajor axis and radius of the planet. Because the ejection is more efficient beyond the snowline (as the Safronov number for a planet of a given mass increases with orbital distance), the majority of the ejected planetesimals are expected to be icy. However,  objects subject to multiple close passages by their host star  could be ejected from closer in \citep{Raymond18}.  In the Solar System, all of the giant planets have $\Theta >$ 1  while the interior terrestrial planets have $\Theta <$ 1. Therefore, the material ejected from the Solar System mostly came from the outer regions, with Jupiter and Neptune responsible  for most of the ejections and the population of the Oort Cloud and Kuiper belt. It has been argued that if interstellar planetesimals come from a ubiquitous and isotropic distribution, it is possible that Jupiter and Neptune analogues are typical in extrasolar systems \citep{Laughlin2017}.  

The estimate shown in Table \ref{Table:galactic_pop} by \citep{MoroMartin2018i} labeled {\it singles} corresponds to the contribution to the interstellar object population from the ejection of planetesimals from protoplanetary around single stars and close binaries, as a result of the planetesimal and planet formation process. In this calculation, it is assumed that the disk mass is 1\% that of the stellar mass, that 1\% is in the form of solids, and that most of the solids are ejected.

\subsubsection{Contribution from Circumbinary Disks}

Intriguingly, circumbinary systems can also be a source of ejected material. Importantly, this material would be ejected from much closer into the potential well of the system. This has  implications for the composition of the ejected bodies. In \citep{Jackson2018MNRAS}, it was estimated that approximately $\sim1/3$  of the ejected material would be icy. The remaining fraction would have spent significant time close to the binary stars prior to  ejection, becoming devolatized. This was used initially as an explanation for the lack of outgassing activity in 1I/`Oumuamua \citep{Cuk2017,Jackson2017}. It is intriguing that some interstellar comets may come from circumbinary systems, and be representative of material that formed at closer stellocentric distances. Moreover, \citep{Childs2022} demonstrated that  \textit{misaligned} circumbinary disks are even more efficient at ejecting interstellar comets. 

The estimate shown in Table \ref{Table:galactic_pop} labeled {\it binaries}  \citep{MoroMartin2018i}  corresponds to  interstellar objects ejected from  circumbinary disks. This estimate assumes, based on \citep{Jackson2018MNRAS}, that the mass of the circumbinary disk is 10\% of the binary system, that 10\% of that material migrates due to gas drag and crosses the unstable radius at which point the objects are ejected, and that 1\% of that material is solids.

\subsubsection{Contribution from exo-Oort clouds}

Another potential source of interstellar planetesimals are exo-Oort clouds. These are swarms of planetesimals that, like the Solar System Oort cloud, are thought to be weakly bound to the central star. The detection of these Oort clouds lies beyond our current observational capabilities, even for the Solar System Oort cloud.  Based on the observed  flux of LPCs in the Solar System, it is estimated that the Oort Cloud harbors $\sim$10$^{12}$ objects larger than 2.3 km (\citep{Brasser13} and references therein).  Oort clouds are thought to be a source of interstellar planetesimals because, as the central star reaches the end of its lifetime, it loses mass in its transition to a white dwarf. This mass loss and the subsequent winds can release the weakly-bound Oort cloud objects into the interstellar medium \citep{Hansen2017,Rafikov2018b,Katz2018}. Exo-Oort cloud objects can also be released via close encounters with other stars \citep{Pfalzner21} or the galactic tide \citep{Veras2011MNRAS,Veras2012MNRAS,Veras2014MNRAS}.

The estimate shown in Table \ref{Table:galactic_pop}  labeled {\it Oort cloud} \citep{Moro2019exOC}  corresponds to the contribution to the interstellar object population from the release of these putative exo-Oort clouds. Because we lack exo-Oort cloud observations, it is  not clear whether or not typical stars harbor exo-Oort clouds. A study by \citep{Brasser10} found that the Oort cloud formation efficiency is similar at a wide range of Galactocentric distances. However, \citep{Wyatt2017MNRAS} found that the parameter space (in terms of planetary architecture) to form an Oort cloud is quite restricted.   The estimate shown in Table \ref{Table:galactic_pop}   labeled {\it Oort Cloud} \citep{Moro2019exOC} adopts the simplifying assumption that exo-Oort clouds are ubiquitous, that they have a population similar to that of the Oort cloud, but scaled to the stellar mass, 10$^{12}\, {{\it M_{*}} / {M_{\odot}}}$, and that they are located at distances similar to that of the Oort cloud, but scaled to Hill radius of its parent star in the Galactic potential. The calculation further assumes exo-Oort cloud clearing is caused by  post-main sequence mass loss for  stars with 1--8 M$_{\odot}$ and stellar encounters for stars that are still on their main sequence.  It assumes varying ejection efficiencies as a function of the stellocentric distances based on previously published dynamical models \citep{Veras2011MNRAS,Veras2012MNRAS,Veras2014MNRAS}. 

However, these post-main sequence stars are necessarily old, and have been subject to dynamical heating.  Therefore, their ejected objects  are expected to have very large velocities as they enter the Solar System. This contrasts with the low velocity of 1I/`Oumuamua, indicating that this object is unlikely to have originated as ejecta from a post main-sequence system. The same is true for 2I/Borisov, given its hypervolatile composition.

There are many uncertainties in the calculations shown for the different potential sources. However, the population studies generally find that there is a discrepancy between the estimated  number density of interstellar planetesimals and that inferred from the detection of 1I/`Oumuamua. This discrepancy indicates that there is much to learn about the population of interstellar planetesimals and their origin. One solution proposed by \citep{MoroMartin2018i} could be that 1I/`Oumuamua was not representative of an isotropic distribution of interstellar planetesimals. This could be explained if the object originated in a nearby planetary system.  As discussed below, this scenario would be consistent with 1I/`Oumuamua's kinematics.   

   \subsection{Kinematics}\label{subsec:galactic_kinematics}

The incoming velocities  of interstellar objects  provide critical information regarding their ages and galactic histories. However, meaningful interpretations of the kinematics of interstellar objects with regards to their provenance most likely will require a larger statistical sample.

In the Milky Way, gravitational interactions with Giant Molecular Clouds (GMCs) and other substrucutures tends to gradually increase the velocity dispersions of stars \citep{BinneyMerrifield1998,BinneyTremaine2008}. This process is colloquially referred to as dynamical heating, and manifests most obviously in the out of plane z-component of galactic velocities. The thin disk of the Milky Way contains younger stars with lower z-dispersion, while older stars that have experienced more dynamical heating populate the thick disk and have larger  excursions out of the galactic plane. Younger stars like B0 stars have typical dispersions of 10s of km s$^{-1}$, while older stellar populations exhibit higher dispersions  \citep{Holmberg09}. Transient structures like GMCs are much younger and have much lower dispersions $\sim 1$ km s$^{-1}$ \citep{MivilleDeschenes2017}. However, these estimates are complicated by a large systematic uncertainty in the velocity of the Sun with respect to the Local Standard of Rest (LSR) \citep{Schonrich12,Robin17}.

1I/`Oumuamua exhibited a surprisingly low hyperbolic velocity of 26 km s$^{-1}$ when compared to the 15$\pm$2 km s$^{-1}$ velocity of the Sun relative to the LSR  \citep{mamajek2017}. 2I/Borisov exhibited a larger relative velocity of 32 km s$^{-1}$. This is indicative that 2I/Borisov has been subject to substantially more dynamical heating than 1I/`Oumuamua, which is presumably due to an older galactic age. The approximate dynamical ages of these objects have been inferred to be $\tau_s \sim10^8$ yr for 1I/`Oumuamua \citep{mamajek2017, Gaidos2017a,Hallatt2020,Hsieh2021} and for 2I/Borisov $\tau_s \sim10^9$ yr \citep{Hallatt2020}. It should be noted that these ages are only statistical in nature.

Another indicator of a young age for 1I/`Oumuamua is its suggested tumbling state with \citep{Drahus18} hypothesizing that the tumbling state could be related to a collision in its parent system. This would suggest that the object is younger than  $\sim$ 1 Gyr, corresponding to the damping timescale due to stresses and strains  resulting from  complex rotation for an object with 1I/`Oumuamua's inferred properties.  A recent ejection is also consistent with the fact that 1I/`Oumuamua’s surface did not seem to be heavily processed based on its color. As mentioned in subsection \ref{subsec:colors}, unlike the cold classical Kuiper belt objects that are ultra-red because of billions of years of exposure to cosmic rays, plasma and radiation, 1I/`Oumuamua’s surface was not ultra-red. 

\subsection{Tracing Interstellar Objects to Progenitor Systems}

The statistical nature of stellar dynamics makes it  difficult to infer a home system for a single interstellar object based on the galactic kinematics. Moreover, the chaotic nature of the galactic motions and the uncertainty in stellar kinematic measurements complicates this task further. In \citep{mamajek2017}, it was suggested that 1I/`Oumuamua's trajectory was sufficiently distinct from all local stars such that it was not co-moving with any single one and therefore not associated with any local exo-Oort clouds. However, \citep{Gaidos2017a} suggested that  the kinematics of 1I/`Oumuamua pointed to its formation in a protoplanetary disk in the Carina or Columba young stellar associations. 

Rigorous analysis of the galactic history of 1I/`Oumuamua concluded that it likely originated within 1 kpc of the Earth, based on its young dynamical age \citep{Hallatt2020}. While identification of a home system is not possible, they reported that 1I/`Oumuamua likely passed through a large subset of the Carina and Columba moving groups at the time that those groups were forming. A similar set of dynamical simulations performed by \citep{Hsieh2021} were consistent with these results, pointing towards formation within a Giant Molecular Cloud. As for 2I/Borisov, \citep{Hallatt2020} demonstrated that identifying its home system is effectively impossible. 

As pointed out earlier, it is critical to note that if 1I/`Oumuamua had a local origin, the implied  $n \sim$ 0.1 au$^{-3}$ number density based on its detection may not apply to the entire Galaxy. Furthermore, \citep{MoroMartin2018i} argued that if 1I/`Oumuamua was ejected from a young nearby star, the spatial density may be highly anisotropic. This would somewhat resolve the discrepancy between the expected number density of interstellar objects and that inferred from the detection of 1I/`Oumuamua which had assumed an isotropic distribution. 

Other attempts to trace 1I/`Oumuamua to a home system were unsuccessful. In \citep{Dybczynski2017}, the search for previous stellar close encounters with 1I/`Oumuamua resulted in none. They noted a $\sim2.2$ pc close encounter with  the nearby planet-bearing star Gliese 876 \citep{Rivera2010}, but this distance is too far to provide a definitive association. A different study by \citep{Feng2018} broadened this analyses and identified 109 potential close encounters but no definitive home system. Another study by \citep{Zuluaga2017} presented a generalized method to estimate the probability that an interstellar object is associated with a stellar system, a methodology that should be useful when a statistical sample of interstellar object kinematics is available. They noted a slight association with the binary star system HD200325.  Even with  accurate stellar kinematics from the European Space Agency (ESA) mission Gaia \citep{Gaia2016}, \citep{Zhang2017} and \citep{bailer-jones2018} reported unlikely prospects to trace a single interstellar object to a host system. 

\subsection{Size-Frequency Distribution of Interstellar Comets}\label{sizedist}

Constraining the size-frequency distribution of the galactic population of interstellar objects is difficult at the large end of the spectrum because there are only two detections of macroscopic interstellar objects. However, there is a long history of studies focusing on the size-frequency distribution of interstellar dust, showing a well-established power-law. Extrapolations of these distributions are tentatively consistent with the implied occurrence rate of 1I/`Oumuamua \citep{Jewitt2022ARAA}, although caution should be used when extrapolating based on such small number statistics.

 Measurements of interstellar dust date back to the Ulysses and Galileo spacecrafts. Combined cumulative results from both of these spacecrafts were reported in  \citep{Landgraf1998} and individual spacecraft measurements were reported in \citep{Grun2000}, \citep{Grun1997} and \citep{Grun1993}.  A controversial detection of an interstellar meteor with the Arecibo Observatory  was reported in \citep{Mathews1999}. Subsequent radar measurements with the Arecibo UHF (430 MHz) were reported by \citep{Meisel2002a} and \citep{Meisel2002b}. The first radar-based detection of micron-sized hyperbolic meteors with the Advanced Meteor Orbit Radar (AMOR) was attributed to the debris-disk host star $\beta$ Pic \citep{Baggaley1993,Baggaley2000,Taylor1996}.  The lower limit from the Canadian Meteor Orbit Radar (CMOR) from twelve possible interstellar events from 1.5 million measured orbits were reported by \citep{Weryk2004}.

At the larger end of the spectrum, \citep{Hawkes1999} analyzed the extant video data (multi-station photographic and television techniques) to constrain the flux of large interstellar meteoroids. An upper limit using optical data with the Canadian Automated Meteor Observatory (CAMO) was reported  by \citep{Musci12}. An extensive analysis of the  IAU Meteor Data Center photographic database  showed that the vast majority of the apparent hyperbolic meteors were a consequence of measurement error \citep{Hajdukova1994,Hajdukova2002}. 

These data are  broadly compatible with a $r_n^{-3}$ size distribution for interstellar particles, including for 1I/`Oumuamua and free-floating planets \citep{Moro2019exOC,Gould22,Jewitt2022ARAA}. Larger  particles  in principle could produce interstellar meteors. These  would have large excess velocities and  trajectories with eccentricies $e>1$. However, these interstellar meteors are notoriously difficult to identify because planetary perturbations can scatter  particles onto hyperbolic trajectories \citep{Wiegert2014,Higuchi20}. Therefore, extremely precise measurements of the incoming velocity vector would be required to verify a meteor as interstellar \citep{Hajdukova2020}. Recently, there have been claims of interstellar meteors detected in the CNEOS database \citep{PenaAsensio2022,Siraj2022meteor}.   However, it is well established that these observations are susceptible to poor accuracy of data. The reported velocity vectors are most notoriously untrustworthy, which are instrumental in aquiring pre-impact trajectories and meteorite fall positions \citep{Devillepoix2019}. Moreover, the number of apparently hyperbolic events in databases, which have been routinely detected for decades, is used primarily as an estimation of the typical error in the detection method.  The  claims of the interstellar nature of meteorites are largely disproven due  to the poor quality of the data and the nondisclosure of error bars \citep{Vaubaillon2022}. 

 As discussed in subsection \ref{GalacticPop}, the comparison of the expected mass budget of ejected planetesimals from extrasolar planetary systems, based on population studies, to the mass budget of interstellar planetesimals, inferred from the detection of 1I/`Oumuamua, shows a significant discrepancy \citep{MoroMartin2018i}. A critical assumption in the calculation of the latter is the albedo and size of 1I/`Oumuamua, and the expected size distribution of interstellar objects. Given the large uncertainties in the latter, the size distribution that is adopted is based on the small body population of the Solar System. This allows for a wide range of possible values that define a broken power-law \citep{MoroMartin2018i}.  Even when adopting this wide range of values, the discrepancy in the mass budget is still present. 

\subsection{Captured Interstellar Objects in the Solar System?}

It has been hypothesized that the Solar System could capture unbound objects from the Galaxy via its gravitational perturbations. This process requires some external force to modify the hyperbolic  trajectory of an interstellar object to make it  bound by the Solar System. This capture process could be driven by external stellar flybys of the tidal galactic gravitational field. Alternatively, it may be driven by more localized   perturbations from a Solar System planet. It has also been hypothesized that some of the Oort cloud comets in the Solar System were captured from other Oort clouds around other stars. This would require the tidal galactic field to impart torques on passing objects. This process may have been more efficient  during the earliest phases of the  formation of the Solar System, when it was in a birth cluster containing many stars \citep{Levison10}. More recent calculations have estimated that the Solar System harbors $\sim10^7$ trapped 1I/'Oumuamua-size interstellar objects within $\sim$ 20 000 au \citep{Penarrubia22}.

The interstellar objects that pass  interior to Neptune could in theory come close to one of the planets and be captured into the Solar System. Since the discovery of 1I/`Oumuamua, several studies have performed detailed calculations to quantify the effectiveness of this process  \citep{Napier2021,Dehnen2022I}. By in large these studies agree  that it is a relatively ineffective process, and only  $\simeq10^{-9}\,M_\oplus$ of interstellar material should be present in the Solar System  \citep{Napier2021b}. With a different set of assumptions, \citep{Dehnen2022II} estimated  that $\sim 8$ captured interstellar objects would exist within 5 au of the Sun. 

There was a claim that Centaurs and Trojans with high inclination, retrograde orbits were captured interstellar objects \citep{Namouni20}. However, this was quickly disproved, when \citep{Morbidelli2020} pointed out that these objects are more likely to come from transient capture of Oort cloud comets. It therefore seems unlikely that captured interstellar objects will be detected in the Solar System. 

\subsection{Interstellar Planetesimals Seeding Planet Formation}

A long standing obstacle in the theory of planetary formation is the so-called  ``meter barrier". The formation of cm-sized particles is efficient and has been well studied \citep{Weidenschilling1977, Weidenschilling1980}. At the meter size, however, collisions between particles have increasing collisional energies and are prone to trigger shattering events \citep{Zsom2010}. Moreover, these larger objects are subject to rapid inward drift timescales as the protostellar gas provides a drag force which removes angular momentum and causes the particles to quickly inspiral. This process significantly limits the lifetime of growing planetesimals in the disk. Several theoretical advances have been proposed to alleviate this problem: the streaming instability which rapidly agglomerates particles into larger objects on secular timescales \citep{Youdin2005}; the direct formation via gravitational instability \citep{Youdin2002}; the presence of low-velocity collisions in the velocity distribution of the dust particles, favoring growth (\citep{Windmark2012}; the trapping of dust grains in pressure maxima \citep{Johansen2004}; and the effect of low porosity and the electric charging of dust aggregates to favor sticking over fragmentation \citep{Okuzumi2009,Okuzumi2009elecric}. 

The discovery of the first interstellar objects led to an additional proposal to alleviate  the ``meter barrier”: the capture of interstellar objects by gas drag into star- and planet-formation environments. This could occur during an early stage, when the objects are trapped into Giant Molecular Clouds. The captured objects tracing the collapse of the prestellar core may subsequently become incorporated into fragmenting star- and planet-forming regions \citep{Pfalzner2019,Pfalzner2021}. Moreover,  interstellar objects may also be captured by already formed protoplanetary disks \citep{Grishin2019MNRAS}. These ideas were further explored by \citep{Moromartin2022} who considered the efficiency of capturing and incorporating interstellar planetesimals into star- and planet-forming regions, as a function of their  velocity dispersion, size distribution and number density.  They found that, when assuming a background number density of 0.2 au$^{-3}$ (derived from \citep{Do2018}), a velocity dispersion of 30 km s$^{-1}$ (characteristic of the young stars in the Galaxy from where interstellar objects likely originated), and an equilibrium size distribution of $q'$ = 3.5, the number of interstellar objects captured by a Giant Molecular Cloud and expected to be incorporated into each protoplanetary disk during their formation would be $\mathcal{O}$(10$^{9}$) (50 cm–5 m), $\mathcal{O}$(10$^{5}$) (5–50 m), $\mathcal{O}$(10$^{2}$) (50–500 m), $\mathcal{O}$(10$^{-2}$) (500 m–5 km). During a later stage, when the protoplanetary disk is formed, the number of interstellar objects that  could  be captured from the interstellar medium during its lifetime would be $\mathcal{O}$(10$^{12}$) (50 cm–5 m), $\mathcal{O}$(10$^{8}$) (5–50 m), $\mathcal{O}$(10$^{5}$) (50–500 m), $\mathcal{O}$(10$^{1}$) (500 m–5 km). In an open cluster where $\sim$1\% of stars have undergone planet formation, these values would increase by a factor of $\mathcal{O}$(10$^{2}$--10$^{3}$). These trapped interstellar objects might be large enough to rapidly grow into larger planetesimals via the direct accretion of the subcm-sized dust grains in the protoplanetary disk before they drift inwards due to gas drag. This provides a promising alternative avenue to overcome the meter barrier, with interstellar planetsimals acting as ``seeds” for planet formation.

The estimates above by \citep{Moromartin2022} are very uncertain because, as discussed in subsections \ref{subsec:galactic_number} and \ref{sizedist}, the number density and velocity and size distribution of interstellar objects are uncertain.  These estimates will be substantially refined as the population of interstellar objects becomes better characterized. However, these preliminary studies have shown that as the number density of interstellar planetesimals in the Galaxy increases with time, their trapping in star- and planet-formation environments may be significant. Therefore, future star- and planet-formation models should take into account the presence of this population of captured objects to assess if it can significantly influence planet formation,  particularly in cluster environments.

\section{Future Prospects}

\subsection{Orbits of Interstellar Objects in the Solar System}
There has been significant interest in identifying the future prospects for the detection and characterization of interstellar objects. As discussed previously (subsection \ref{subsec:galactic_number}), the discovery of 1I/`Oumuamua, if representative of an isotropically distributed population, might imply that a galactic population of similar objects exists with spatial number density of  $n\sim 0.1~{\rm au}^{-3}$. There has therefore been significant efforts invested into identifying the expected orbital distributions of interstellar objects that travel through the Solar System.

The kinematic distribution of local stars in the local stellar neighborhood is well measured (see subsection \ref{subsec:galactic_kinematics}). Moreover, interstellar objects should represent a better realization of the kinematics of the gravitational potential of the Milky Way since there are estimated to be orders of magnitude more of them than stars in the Galaxy.   Assuming that the incoming interstellar objects trace this velocity distribution,  the distribution of orbits expected for the interstellar objects that pass through the Solar System can be calculated. 

\begin{figure}%[h]
\begin{center}
       \includegraphics[scale=0.42,angle=0]{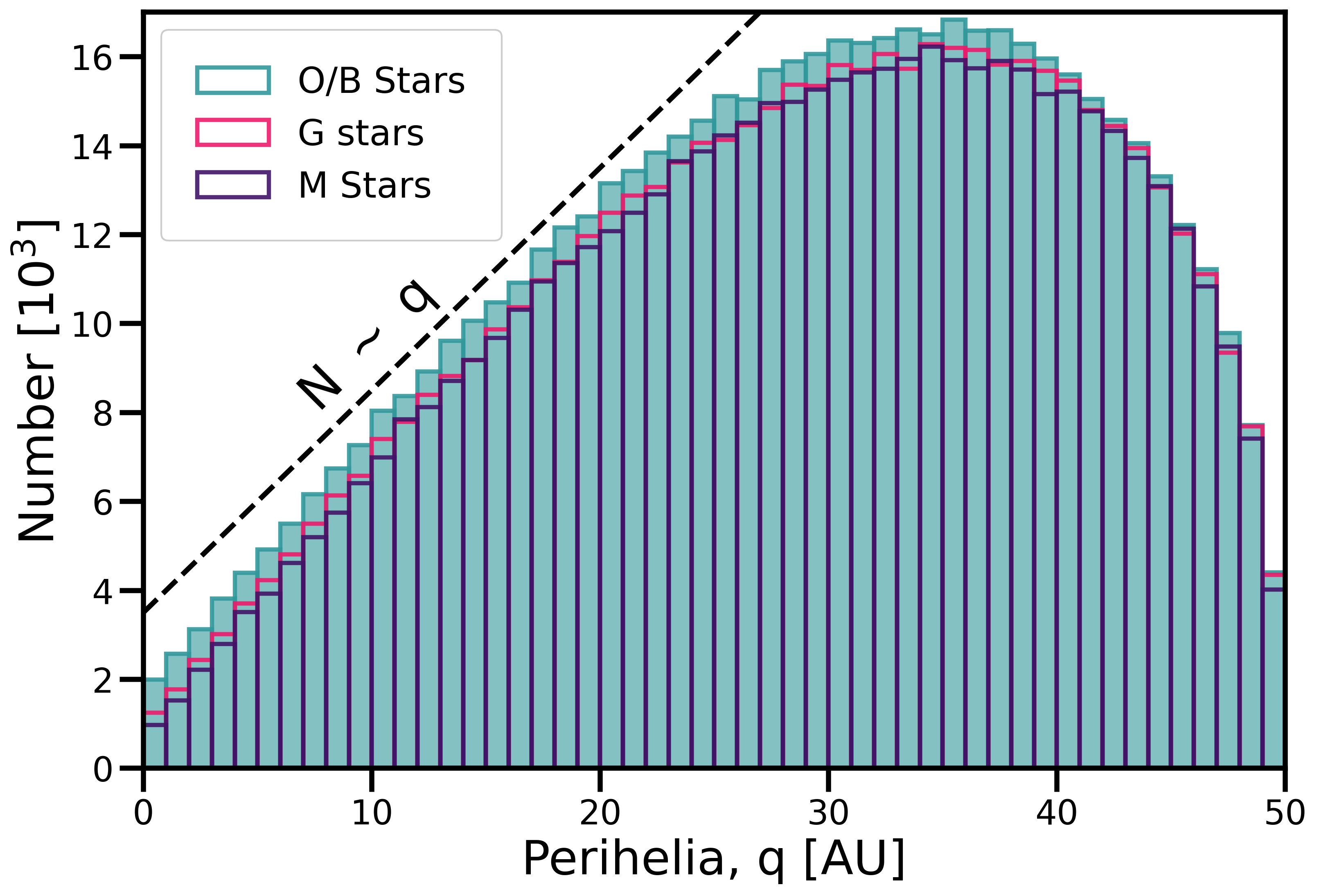}
    \caption{ Predicted perihelia distribution of interstellar objects from the probabilistic method generated by \citep{Marceta2023}. Distributions are shown assuming incoming kinematics of various stellar populations. Reproduced with permission from \citep{Marceta2023}.}\label{fig:orbits} 
\end{center}
\end{figure}

This is straightforward to implement numerically via a  Monte Carlo method. This was performed  by \citep{Cook2016} and \cite{Engelhardt2014} prior to the detection of 1I/'Oumuamua. After its discovery, with updated number densities, these simulations were re-performed and modified by \cite{Seligman2018} and \cite{Hoover2022}. With an elegant mathematical analysis, \cite{Marceta2023} provided an entirely analytic method to calculate the distribution of interstellar object orbits. They showed comparisons of their analytic  method (which they label the probabilistic method) -- with the numerical method previously performed (that they refer to as the dynamical method), demonstrating that the probabilistic method is  orders of magnitude more computationally efficient and also more accurate. The resulting distribution of perihelia is shown in Figure \ref{fig:orbits}, for interstellar comet populations with galactic velocity dispersions of various stellar populations. In all cases, the number of interstellar objects passing through the Solar System varies linearly with perihelia. 

\subsection{Ground Based Prospects}

The forthcoming Rubin Observatory Legacy Survey of Space and Time (LSST)  \citep{jones2009lsst,Ivezic2019} will be efficient at detecting transient objects \citep{solontoi2011comet,Veres2017,veres2017b,Jones2018}. Literature estimates based on survey detection efficiency and criteria and number densities estimate that the survey will detect $\sim$1-3 1I/`Oumuamua-like interstellar object per year \citep{Moro2009,Engelhardt2014,Cook2016,Trilling2017,Seligman2018,Hoover2022,Marceta2023}. 

The most recent study of the of the expected distribution of interstellar objects by \citep{Hoover2022} included a calculation of the distribution of orbital elements and close approaches to the Earth of interstellar objects that will be detected by the LSST. They incorporated a set of LSST detection criteria in this analysis that required that the elongation angle and phase be such that the the LSST would be able to detect the object. The results of this simulation are shown in Table \ref{percent_table}, and show that the LSST should detect 1-2 interstellar objects per year.

\subsection{Space Based Prospects}

The forthcoming NEO Surveyor \citep{Mainzer2015} may also detect interstellar objects. NEO Surveyor has  a 50 cm diameter telescope and will be located interior to Earth's orbit at  L1. The mission has been optimized to  find  small bodies out to Jupiter that could be potentially hazardous. Moreover, the mission will provide thermal infrared measurements of these objects.  NEO Surveyor and possibly \textit{JWST} will be able to characterize both the sizes and albedos of these objects.

Recently, ESA selected the \textit{Comet Interceptor} \citep{jones2019} which will fly  in 2029. \textit{Comet Interceptor} has a low $\Delta V \sim$ 1 km s$^{-1}$  budget, and is designed to rendezvous with a LPC.  If an interstellar object is detected with a serendipitously fortuitous orbit, then it is possible that \textit{Comet Interceptor} will reroute and rendezvous with it. For example, \citep{Seligman2018} showed that an impactor mission to 1I/`Oumuamua sent from Earth would have been achievable with a $\Delta V \sim$ 4 km s$^{-1}$ impulse given an earlier detection and sufficient lead time. About 10\% to 30\% of the interstellar interlopers to be detected by the LSST (Table \ref{percent_table}) should be reachable by a mission with $\Delta$V $<$ 15 km s$^{-1}$ \citep{Hoover2022}. There has been ample work done to design such a dedicated mission to an interstellar object \citep{Hein2017,Seligman2018,Meech2019whitepaper,Castillo-Rogez2019,jones2019,Hibberd2020,Linares2020,Donitz2021,Sanchez2021,Meech2021,Hibberd2022,Moore2021whitepaper,Moore2021,Hein2022,Mages2022,Miller2022,Garber2022}, and it seems likely that a handful of reachable targets will be detected. 

The study of interstellar objects, still in its infancy, will experience a revolution with upcoming ground-based and space-based observations. These will not only  help us understand the origin(s) of this new component of the interstellar medium and the clues it unveils regarding planet formation, but will also allow for the  extraordinary opportunity to study a fragment from another planetary system at close-range. 

\begin{table}
\caption{The  frequency that interstellar objects are detectable by the LSST and reachable with a range of $\Delta$V criteria. These calculations   assume that each interstellar object has the same absolute magnitude as that of 1I/`Oumuamua. Reproduced from \citep{Hoover2022}.  }
    \centering
 
    %\vspace{0.5em}
    
    \begin{tabular}{c c c c}
    
    Criterion & Percent\,\,\,& Conservative Rate\,\, &Optimistic Rate\\
     & &  per Year& per Year\\
    \hline
    $m\leq 24$ & $\sim$7.0\% & $\sim 2.3$ & $\sim 4.6$\\
    
    Detectable with the LSST & $\sim$2.8\% & $\sim 0.9$ & $\sim 1.9$\\
    
    Detectable, $\Delta V < 30$ km/s & $\sim$1.1\% & $\sim 0.35$ & $\sim 0.7$\\
    
    Detectable, $\Delta V < 15$ km/s & $\sim$0.4\% & $\sim 0.1$ & $\sim 0.3$\\
    
    Detectable, $\Delta V < 2$ km/s & 0.002\% & $\sim 7\times10^{-4}$ & $\sim 0.001$\\\hline
    \end{tabular}
    
    \label{percent_table}
\end{table}

\section*{Acknowledgement(s)}
We thank Du{\v{s}}an Mar{\v{c}}eta for providing simulation results for Figure 10. We thank Davide Farnocchia and Marco Micheli for  detailed reviews of the manuscript prior to submission. We thank Jennifer Bergner, Aster Taylor, Lisa Kaltenegger, Dave Jewitt, Adina Feinstein, Fred Adams, Dong Lai, Nikole Lewis, Greg Laughlin and Juliette Becker for useful suggestions and conversations.  

%An unnumbered section, e.g.\ \verb"\section*{Acknowledgements}", may be used for thanks, etc.\ if required and included \emph{in the non-anonymous version} before any Notes or References.

%\section*{Disclosure statement}

%An unnumbered section, e.g.\ \verb"\section*{Disclosure statement}", may be used to declare any potential conflict of interest and included \emph{in the non-anonymous version} before any Notes or References, after any Acknowledgements and before any Funding information.

\section*{Funding}

%An unnumbered section, e.g.\ \verb"\section*{Funding}", may be used for grant details, etc.\ if required and included \emph{in the non-anonymous version} before any Notes or References.
DZS acknowledges financial support from the National Science Foundation  Grant No. AST-2107796, NASA Grant No. 80NSSC19K0444 and NASA Contract  NNX17AL71A from the NASA Goddard Spaceflight Center. 

%\section*{Notes on contributor(s)}

%An unnumbered section, e.g.\ \verb"\section*{Notes on contributors}", may be included \emph{in the non-anonymous version} if required. A photograph may be added if requested.

\bibliographystyle{tfnlm}
\bibliography{interactnlmsample}

\end{document}